\documentclass[useAMS, usenatbib]{mnras}
\usepackage{hyperref}
\usepackage{xspace}
\usepackage{amsmath}
\usepackage{graphicx}
\usepackage{float}
\usepackage{caption}
\floatstyle{plain}
\restylefloat{figure}

\usepackage{tabularx}
\usepackage{cleveref}
\crefname{section}{\S}{\S\S}
\Crefname{section}{\S}{\S\S}

\newcommand{\ltsima}{\mbox{$\; \buildrel < \over \sim \;$}}
\def\simlt{\lower.5ex\hbox{\ltsima}}
\def\gtsima{$\; \buildrel > \over \sim \;$}
\def\simgt{\lower.5ex\hbox{\gtsima}}

\def\ie{{\it i.e.}}

\def\ltsima{$\; \buildrel < \over \sim \;$}
\def\simlt{\lower.5ex\hbox{\ltsima}}
\def\gtsima{$\; \buildrel > \over \sim \;$}
\def\simgt{\lower.5ex\hbox{\gtsima}}

\def\H2{H$_2$\xspace}

\def\h2{H$_2$\xspace}
\def\ion#1#2{\text{#1\,\sc #2}}
\def\HI{{\ion{H}{i} }}
\def\HII{{\ion{H}{ii} }}
\def\GI{{\ion{He}{i} }}
\def\GII{{\ion{He}{ii} }}
\def\GIII{{\ion{He}{iii} }}

\def\pop3{Population~III\xspace}

\def\hide#1{}

\pdfoutput=1

\title[ARC: Adaptive Ray-tracing with CUDA]{ARC: Adaptive Ray-tracing with CUDA, a New Ray Tracing Code for Parallel GPUs}

\author[Hartley \& Ricotti]{Blake Hartley$^{1}$ and Massimo Ricotti$^{1}$\\
$^1$Department of Astronomy, University of Maryland, College Park, MD 20742, USA
\thanks{bth@astro.umd.edu, ricotti@astro.umd.edu}
}
\begin{document}
\maketitle

\begin{abstract}
We present the methodology of a photon-conserving, spatially-adaptive, ray-tracing radiative transfer algorithm, designed to run on multiple parallel Graphic Processing Units (GPUs). Each GPU has thousands computing cores, making them ideally suited to the task of tracing independent rays. This ray-tracing implementation has speed competitive with approximate momentum methods, even with thousands of ionization sources, without sacrificing accuracy and resolution.
Here, we validate our implementation with the selection of tests presented in the "cosmological radiative transfer codes comparison project," to demonstrate the correct behavior of the code. We also present a selection of benchmarks to demonstrate the performance and computational scaling of the code. As expected, our method scales linearly with the number of sources and with the square of the dimension of the 3D computational grid. Our current implementation is scalable to an arbitrary number of nodes possessing GPUs, but is limited to a uniform resolution 3D grid. Cosmological simulations of reionization with tens of thousands of radiation sources and intergalactic volumes sampled with 1024$^3$ grid points take about 30 days on 64 GPUs to reach complete reionization.
\end{abstract}

\section{Introduction}\label{sec:intro}

The propagation of ionizing and dissociating radiation from stars and black holes and its effect on the interstellar medium (ISM) and the intergalactic medium (IGM), is one of the most fundamental and computationally difficult problems in theoretical astrophysics. A number of schemes have been implemented for tackling radiative transfer. As the computational power available for astrophysical simulations has increased over the past few decades, the full seven dimensional (three spatial, two angular, one frequency, one time) radiative transfer problem has been solved in earnest. There are multiple popular methods for approaching this problem, including:

\begin{enumerate}
\item Moment Methods: The first three moments of the radiation intensity, which are the energy density, flux, and radiation pressure, are tracked by the simulation \citep{1970MNRAS.149...65A, 1998MmSAI..69..455N, 1992ApJS...80..819S}. Simulations of this type have been implemented with short characteristics \citep{1992ApJS...80..819S}, long characteristic \citep{2009MNRAS.393.1090F}, utilizing the optically thin variable Eddington Tensor method \citep{2001NewA....6..437G, RGS2002a, RGS2002b, 2009MNRAS.396.1383P}, and with a two moment model using a closure relation \citep{2007A&A...464..429G, 2008MNRAS.387..295A}. These methods have the advantage of being fast, with computation times that don't depend on the number of radiation sources. However, these methods are fundamentally diffusive, so that radiation will in some cases flow around occluding regions in non-physical ways, resulting in incorrect behavior for shadows. In addition, since the radiation is treated as a photon fluid with characteristic speed equal to the speed of light, the condition for stability of the integration typically requires taking very small time step (similarly to the CFL condition). A reduced speed of light approximation is often adopted in order to be able to take larger timesteps \citep{Deparisetal2018}.

\item Ray-tracing Methods: Radiation from a point source is approximated as a collection of linear rays propagating away from the point source. Along the rays the heating/ionization of the gas and extinction of the radiation are tracked through a grid \citep{1999ApJ...523...66A, 1999MNRAS.309..287R, 2001MNRAS.324..381C, 2001NewA....6..359S, 2006ApJ...639..621A, 2006NewA...11..374M, 2006A&A...452..907R, 2006ApJS..162..281W, 2007ApJ...671..518K, 2007ApJ...671....1T, 2010A&A...515A..79P} or collection of particles \citep{2006PASJ...58..445S, 2007ApJ...665...85J, 2008MNRAS.386.1931A, 2008MNRAS.389..651P, 2011MNRAS.412.1943P, 2009MNRAS.395.1280H}. These methods are computationally more expensive, as the amount of computation scales linearly with the number of sources and the number of grid points or gas particles. The results, however, are a better approximation of the true solution.

\item Non-simulation methods, such as the excursion set formalism \citep[inside-out reionization, see ][]{Furlanetto:2004, 2007MNRAS.380L..30A}, inhomogeneous reionization models \citep[outside-in reionization, see ][]{Miralda-Escude:2000,Wyithe:2003,Choudhury:2005}, and semi-analytic inhomogeneous reionization methods \citep{mitra:2015}. These methods require much fewer computational resources than simulation methods, allowing for the exploration of a much larger range of simulation parameters. However, they lack the accuracy of computational methods.
\end{enumerate}

Our approach is of the second type, based on the photon-conserving, spatially-adaptive ray-tracing method originally presented in \cite{2002MNRAS.330L..53A}, but designed for use on parallel Graphics Processing Units (GPUs). GPUs have slower clock speeds than CPUs of the same generation, but they possess thousands of parallel cores which operate independently of each other. The cost of calculations for each individual ray is computationally small and independent of other rays, which makes GPUs ideally suited for the task. Despite the large ongoing effort to model the epoch of reionization (EOR) by various teams, our proposed approach is complementary to previous and ongoing investigations, and has new and unexplored aspects to it as explained below. We are writing our numerical code with the future goal of incorporating it in widely used hydrodynamical cosmological codes. At the moment, the focus of our effort is on ray-tracing radiative transfer, meaning we limit ourselves to simulations with fixed grid (\ie, the adaptive mesh refinement (AMR) grid structure is not implemented at this point) and no coupling to hydrodynamics, thus making this code a purely post-processing method on previously run cosmological simulations. This approximation is not poor when focusing on simulations of cosmological reionization on scales much larger than galaxy-halo scales, which are the focus of our test runs and future science investigations.

This paper is organized as follows. In Section~\ref{sec:methodology} we describe the underlying physics and computational implementation used by our code. In Section~\ref{sec:tests} we present the results produced by our code when running the battery of tests laid out in the radiative transfer codes comparison project \citep{2006MNRAS.371.1057I}. We also present a selection of benchmarks demonstrating the speed of the code in Section~\ref{sec:bench}. Finally, in Section~\ref{sec:summary} we summarize the methods and results presented in this paper.

\section{Methodology}\label{sec:methodology}

Our code utilizes the power of GPU processing to attack the computationally intensive problem of adaptive ray-tracing radiative transfer. The tracing of a single ray of photons is a very simple task, but a ray tracing radiative transfer simulation requires the tracking of a large number of such rays, making the task well suited to GPU computing. Classically, memory sharing between GPUs is only possible if those GPUs are all located on the same node of a computer. This limits the scale of a purely GPU algorithm, and lead us to add MPI parallelization. MPI parallelization allows us to share the results of GPU calculations between GPUs on different nodes, thus allowing for a highly scalable code and also overcoming the problems associated with the limited memory of a single GPU. In addition to the relatively straightforward parallelization of radiation from different sources, we break up the computational volume and therefore the ray tracing and the ionization/heating calculations into equal portions and distribute them among each node of the process using MPI. There, the GPU algorithm performs the calculations, sharing ray data between nodes using MPI as necessary. The results of the calculations are then consolidated using MPI and the next time step can be taken. An earlier and simpler version of the code shared copies of the whole the grid data among nodes and distributed only the work from different ionization sources using MPI. This way, there was no need for ray data sharing among nodes. However, this method was severely limited by the available memory of each GPU. With current GPUs, the maximum resolution of reionization simulations we could run without breaking up the volume was between 256$^3$ and 512$^3$ grid points. The current code is indeed faster and not limited by these memory restrictions.

\subsection{Program Design}

Each GPU in our code calculates the radiation field produced by sources within the sub-volume assigned to the node which hosts that GPU. In addition each sun-volume can share the work with multiple GPUs. The direction of the rays is assigned using the HEALPix \citep{2005ApJ...622..759G} scheme to distribute rays in all directions. The HEALPix scheme assigns to each ray equal solid angle by dividing the sphere around each sources in equal areas. In order to maintain constant spatial resolution as the rays travel further from the source, each subsequent level of the scheme breaks this area into four sub-areas, allowing the algorithm to easily keep track of the splitting of rays. We initialize an array of photon packets for each ray in the initial HEALPix array (typically HEALPix level 1, or 48 photon packets, or rays). Each photon packet traced by our code contains the ID of the ionizing source which emitted it, the unique HEALPix PID which determines the direction of the photon's motion, the distance each photon has traveled, and the optical depth of each frequency bin along the photons path. The position of the ionizing source is stored in the GPU's shared memory, allowing for a reduction of the required per-photon packet memory.
Once the calculation of the radiation fields by the GPUs is complete, we combine the radiation fields from other sources distributed on different nodes linearly using MPI. We then divide and distribute all these fields to the processes to evolve the gas fields over a single time step. The algorithm's overall structure is described in steps below.
\begin{enumerate}

\item Initialize radiation field arrays to zero at the start of a new time step.
\item Use MPI to divide distribute photon source data and source array to all nodes of the process.
\item Loop through each photon source one at a time using the GPU kernel as follows:
\begin{enumerate}
\item Initialize all photon packets to radius zero and zero optical depth in all frequency bins.
\item Trace all photon packets at a single HEALPix level until the ray moves to an adjacent sub-volume, terminates or splits. If the photon moves into an adjacent sub-volume, we store it in a sharing buffer to be sent later. If the photon terminates, set the radius of the packet to zero, signaling that it should be ignored in future calculations. Otherwise, leave the radius and optical depth at the point of splitting unchanged.
\item Create a new array of child rays populated by the unterminated rays from the previous step. The radius of all four child rays is the same as the parent rays, but the newly populated array have directions based on the next level of the HEALPix scheme.
\item If any rays are unterminated, return to step (b).
\item Use MPI to share rays between adjacent sub-volumes, and return to step (a).
\end{enumerate}
\item Use MPI to linearly combine radiation fields produced by all of the kernels, thus giving us the overall radiation field produced by all the sources.
\item Apply the ionization/heating calculation to the current radiation field and gas fields to calculate the changes to the grid over a single time step.
\item Use the execution times of each GPU to redistribute the sources between GPUs handling the same sub-volume to balance the computational load.
\item Return to (i) until the simulation reaches the desired time.
\end{enumerate}

Each node and corresponding GPU loops through this code to advance the simulation. We can also send the same sub-volume to multiple GPU's, allowing us to break a set of sources in the same sub-volume between multiple GPUs. When any timing information about the entire simulation, or outputs of the full grid, are required, MPI is used to consolidate the data from all of the nodes.

In the next subsections we describe in more detail the implementation of the physics and equations solved by each module of the code.

\subsection{Photon Data}\label{photondata}

Photon packets in our code are polychromatic. As each ray is traced through the grid, we calculate the total optical depth due to all atomic species (hydrogen and helium) and their ions, for each frequency bin. Each bin is a single frequency which represents the frequency averaged ionization cross section of all present species. We are able to track an arbitrary number of frequency bins, subject to memory limitations.

\subsection{Grid Data}

The basis of our code is a uniform 3-dimensional cubic grid of $N^3$ cells which store the neutral/ionized fraction, temperature, and photoionization rate of a given set of atomic/molecular species as a function of time. The current version of our simulation tracks the neutral fractions $x_{\HI}$ and $x_{\GI}$ (future versions will include tracking of more species as required). We assume that helium is either neutral or singly ionized over the course of the simulation. Thus, we calculate the electron number density as:

\begin{align*}
n_e=n_b\left[\frac{1-Y_p}{A_{\rm H}}x_{\HII}+\frac{Y_p}{A_{\rm He}}(x_{\GII}+2x_{\GIII})\right],
\end{align*}
where $n_b \equiv n_p+n_n$ is the baryon number density, $Y_p \equiv \rho_{He}/\rho_b$ is the Helium fraction, $A_H=1$ and $A_{\rm He}=4$ are the atomic weights of hydrogen and helium, respectively.

We are able to sub-divide this volume as desired. The first science runs of our code will divide the volume evenly into 8 cubic regions, each of which will be handled by a subset of the available GPUs. The current implementation of the code is limited to a fixed grid; however, the structure of the code would allow for a transition into an adaptive mesh refinement scheme (AMR) should it be required in the future.

\subsection{Photon Transmission}

Ionizing sources within our volume are represented by a set of photon packets, or rays, which are initialized at the location of the source. These rays are traced through the grid of our simulation one step at a time, with each step corresponding to the distance through the volume it takes to reach the next X, Y, or Z cell boundary within the volume. Each ray is initialized with an ionizing photon production rate ($S_{\nu}$ [s$^{-1}$] for each frequency bin ${\nu}$) which corresponds to the luminosity of the source divided between the rays. For each step, we increment the total optical depth along the ray $\tau_{\nu}$ for each frequency bin by $\Delta\tau_{\nu}$, the optical depth of ray within a single simulation cell. We use this optical depth to calculate the fraction of ray's photons absorbed by gas within the cell, and thus calculate the ionization rate of the cell for each frequency bin $\nu$ per absorbing atom:
\begin{align}\label{eq:ion}
s_{\nu}=\frac{S_{\nu}(1-e^{-\Delta\tau_{\nu}})}{V\;n_{\rm abs}},
\end{align}
where $n_{\rm abs}$ is the physical number density of absorbers, and $V$ is the volume of the simulation cell. More specifically, in our code $n_{\rm abs}$ is the neutral number density $n_{SI}$ of a species $S$, where $S$ refers to hydrogen and helium. The flux along the ray in each frequency bin is reduced by this amount, so that this method conserves the total number of photons. Each ray in the simulation is traced by a single CUDA core, with the work distributed as evenly as possible by the CUDA kernel. Each GPU contains all of the grid data necessary to trace the rays, and the ionization rate at a given cell in the grid is altered "atomically" by each core, eliminating the chance for interference between simultaneous processes. 

The directions of the rays are chosen according to the HEALPix scheme \citep{2005ApJ...622..759G}, which assures that each ray is assigned an equal solid angle relative to the source. This means that we can assign a photon emission rate for each ray by dividing the overall $f_{\rm esc,\nu} S_\nu$ of the source evenly between all rays (where $f_{\rm esc, \nu}$ is the mean ionizing radiation escape fraction from each source), or we can adopt a more sophisticated (and realistic) scheme in which the escape fraction is anisotropic. We can, for instance, obtain the wanted escape fraction by only assigning the mean $S_\nu$ per-ray, to a fraction $f_{\rm esc, \nu}$ of the rays, while completely blocking the radiation escaping from the remaining rays. 

\begin{figure}[t!]
\includegraphics[width=8.0cm]{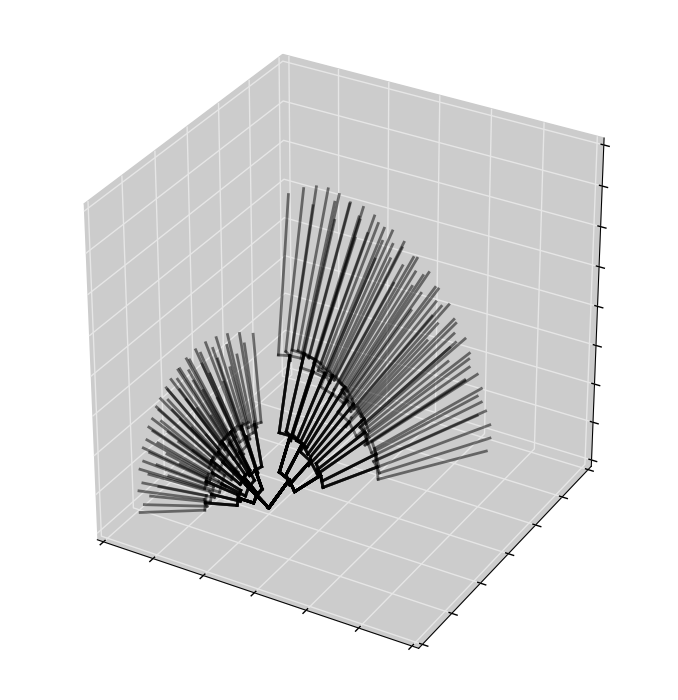}
\caption{Schematic example of the adaptive ray tracing method based on HELPix. As rays propagate outwards, they split to maintain a constant number of rays per unit area crossing the surface of the sphere centered on the source. The left and right trees illustrate the difference between assuming at least 3.0 and 1.0 rays intersect a given simulation cell for the left and right ray trees, respectively.} 
\label{fig:healpixcooling}
\end{figure}

Each ray is traced independently by a CUDA core until it reaches a set distance from the source, at which point the ray is split into four child rays corresponding to the next level of the HEALPix scheme. The distance at which the ray is split is a free parameter which allows us to control how many rays pass through the average simulation cell at a given distance, giving us control over the spacial accuracy of the ray tracing method. 
In Figure~\ref{fig:healpixcooling} we plot a schematic of the adaptive ray tracing process. The two trees represent the progress of a single pixel at the lowest HEALPix level traced out through three branches. The two trees demonstrate the difference between thresholds for ray splitting, with the left tree guaranteeing three times as many pixels through a given cell than the right tree.

It is important to emphasize that, as mentioned above, HEALPix method gives us the ability to control to any desired accuracy the directions into which the ionizing source emits photons, allowing simulations in which ionizing photons leak out of a galaxy anisotropically. Numerical simulations of galaxy formation have shown that "chimneys" of low optical depth through which most ionizing photons escape into the intergalactic voids are a more realistic description of the radiation escaping into the IGM than an overall isotropic attenuation of the emission. We can control these directions on a source by source basis, giving us a versatile tool to study different types of sources in different cosmic environments. This desirable feature of the code is only possible because the ray tracing method is non-diffusive and can capture shadows accurately. The faster "radiation moments" methods discussed in the introduction, are probably too diffusive to allow the implementation of anisotropic emission from the sources.


\subsection{Geometric Correction}\label{sec:correction}

The rays in our code are one-dimensional lines which represent a cone of radiation extending from a point source, or from a splitting ray, with a fixed solid angle. The intersection of the ray volume with a given grid cell is a complicated geometric shape whose volume is impratical to calculate exactly. The ray tracing method approximates this volume as a truncated cone created by the segment of the ray whose length is the distance between the points where the ray enters and exits the grid cell. Thus, when the ray remains closer to the edge of the grid cell than the radius of the ray (along the length of the ray), a portion of the ray remains entirely outside of the grid cell. We make a first order correction to this problem following the method presented in \cite{2011MNRAS.414.3458W}. We let $L_{\rm Pix}$ be the width of the pixel and let $D_{\rm edge}$ be the distance from the midpoint of the segment of the ray within the cell to the nearest edge of the cell. We reduce the ionization within the cell by a factor $f_c$ defined by:
\begin{align}\label{eq:correction}
f_c=
\begin{cases} 
\frac{1}{2} + \frac{D_{\rm edge}}{L_{\rm Pix}} & D_{\rm edge}< L_{\rm Pix}/2, \\
1 & D_{\rm edge}\geq L_{\rm Pix}/2.
\end{cases}
\end{align}
This correction factor is slightly different from the one used in \cite{2011MNRAS.414.3458W} in which $(1/2+D_{\rm edge}/L_{\rm Pix})$ is squared. We remove the square from the first expression, so that $f_c=1/2$ when $D_{\rm edge}=0$, which physically corresponds to half of the ray lying within the grid cell when the ray travels along the edge of the cell. However, this correction factor only serves to reduce the ionization rate in a given cell, and thus has the effect of reducing the global ionization rate in the simulation and breaks photon conservation, that is one of the most desirable properties of the method. We find that adopting this correction makes a non-negligible reduction to the volume filling fraction in the simulation, which is more pronounced at the points where the rays split (as may be seen in the radial profiles of \cite{2011MNRAS.414.3458W} Figure 5 and Figure 6).

We decided to keep the correction factor as it mitigates small spatial artifacts in azimuthal directions (see Section~\ref{ssec:methodtest}), but we also compensate in order to maintain photon conservation by adding the ionization removed by the correction factor to the nearest cell using a secondary correction factor:
\begin{align}\label{eq:correction2}
f'_c=(1-f_c)\frac{x_{\rm abs}}{x'_{\rm abs}},
\end{align}
where $x_{\rm abs}$ and $x'_{\rm abs}$ are the absorber fraction in the cell intersected by the ray, and in the nearest adjacent cell to the ray, respectively.
Here we have multiplied by the ratio of the density of absorbers between the cells, as the ionization rate in Equation~\ref{eq:ion} is an overall absorption rate, and needs to be corrected in case the  the densities of absorbers vary between the cells. This second correction factor is another slight difference with respect to the method used in \cite{2011MNRAS.414.3458W}, but we find it very beneficial. The combination of these correction factors corrects for geometric artifacts while maintaining photon conservation and the correct volume filling factor of the \HII regions (see Section~\ref{ssec:methodtest}). This secondary correction relies on the regularity of the Cartesian grid, and in the case of non-regular grids or AMR it would need to be generalized or removed.

\subsection{Optically Thin Approximation}

In regions of low neutral fraction surrounding active sources of ionization, the ray tracing process tracks an optical depth which is almost unchanging over the grid cells. This allows us to implement a procedure for calculating the ionization within a certain volume without tracing rays, and thus start the ray tracing process at a greater distance and save significant computation time. Our optically thin approximation proceeds as follows:

\begin{enumerate}
\item To each ionizing source within the simulation we assign a radius $R_{\tau<0.1}$, which we initially set to zero.

\item As rays are traced from the ionizing source, we set $R_{\tau<0.1}$ to be the minimum of all the ray lengths for which $\tau_{\nu}<0.1$ in the softest frequency band.

\item When $R_{\tau<0.1} > 0$, we assume all cells within a distance of $R_{\tau<0.1}$ of the ionizing source to be directly exposed to the ionizing source without absorption. We approximate the ionization rate as:
\begin{align*}
s_{\nu}\approx\frac{S_{\nu}x_{\rm abs}\sigma_{\rm abs}(\nu)}{4\pi R^2},
\end{align*}
which is the approximation of Equation \ref{eq:ion} for small $\tau_{\nu}$ and replacing the dilution of spreading rays with a $4\pi R^2$ term.

\item We initialize rays at a radius of $R_{\tau>0.1}$. This allows us to skip ray tracing within the optically thin region.
\end{enumerate}

This approximation is trivial at early times in simulation, when $R_{\tau>0.1}$ is small for most sources, but the amount of calculation is then also small. However, when the average neutral fraction of the simulation decreases, this approximation becomes more and more efficient, saving significant computational resources. Beginning the rays outside of these these highly ionized regions decreases the ray tracing computation time by a factor of the order of the mean neutral fraction in the simulation box, while the calculation itself is orders of magnitude faster than the ray tracing module.

\subsection{Ionization Calculation}

Each GPU in our code loops through all sources it is assigned using the procedure detailed in the previous section. Once all the GPUs have completed their assigned work, they return the radiation array which represents the ionization rate due to all of the sources assigned to that GPU. The host node then sums these arrays using MPI, giving us the overall ionization rate at each point in the grid due to all sources in the simulation.

Once the ionization rate, $\Gamma_{\nu}$, is calculated, we compute the change in the neutral density, $n_{SI}$ of species $S$ of the gas according to the differential equation:
\begin{align}
\dot{n}_{SI}=-n_{SI}\sum_{\nu}s_{\nu}-C_S(T) n_e n_{SI}+\alpha_S(T)n_e n_{SII},\label{eq:diffeq}
\end{align}
where $C_S$ is the collisional ionization rate, $\alpha_S(T)$ is the recombination rate, $n_e$ is the electron number density and $n_{SII}$ is the ionized number density of species $S$.

The ray tracing algorithm is computationally the most expensive part of the simulation, so we have designed the code to limit the number of calls to the ray tracer as much as possible. We solve the non-equilibrium chemistry and energy equations for all species under consideration sub-cycling between the radiative transfer time steps \citep{1997NewA....2..209A}.
The non-equilibirum chemistry equations are stiff ordinary differential equations (ODEs). Thus, for individual steps in the sub-cycle, we tested several integration methods, including predictor-corrector methods, Runga-Kutta methods, semi-implicit methods, backwards difference, etc. We found that the backwards difference formula (BDF) gave the best combination of speed, accuracy, and computational stability in our tests, in agreement with the results presented in \citep{1997NewA....2..209A}.
Effective use of the BDF requires writing the non-equilibrium chemistry equations in the form:
\begin{align}\label{eq:anninos}
\dot{n}_{SI} = D - C\cdot n_{SI},
\end{align}
where $D$ represents source terms which do not depend on $n_{SI}$ (to the first order) and $C$ represents sink terms which are linear in $n_{SI}$ (to the first order). The photon conserving method we use (Equation~\ref{eq:ion}) calculates the photo ionization rate per absorbing atom within a cell at a given neutral fraction.
In order to use Equation~\ref{eq:anninos}, we must assume that the photo-ionization ($s_\nu$) and heating per absorber within the cell, which depend on $(1-e^{-\Delta\tau})/n_{SI}$, is either independent of the neutral fraction or is $\propto 1/n_{SI}$, so that $n_{SI}s_\nu \sim {\rm const.}$ Since $\Delta\tau$ is proportional to $n_{\rm SI}$, we are effectively treating single cells as having either $\Delta \tau \ll 1$ ({\it i.e.}, $1-e^{-\Delta\tau}\propto\Delta \tau\propto n_{\rm SI}$) or $\Delta \tau \gg 1$, over a set of sub-cycles. We note that where this approximation is less accurate, the cells ionize faster than in the accurate case, becoming rapidly optically thin and transitioning into the more accurate regime. We also have the option of treating the cells as optically thick in cases of particularly high density. However, in the bulk of the cases we consider, high density cells at the ionization front are pre-ionized by photons with a long mean-free-path because of a spectrum which has been hardened by absorption along the ray. In summary, the optically thin approximation is almost always sufficiently accurate.

We choose sub-cycle time steps for the ionization/energy ODEs to limit the fractional change in any quantity which our equations is tracking, including the energy of the cell. Thus:
\begin{align*}
dt=\min\left(\frac{\epsilon E}{|dE/dt|},\frac{\epsilon n_{\HI}}{|dn_{\HI}/dt|}\right).
\end{align*}

The choice of $\epsilon$ is free; we find that a choice of $\epsilon = 0.1$ following the convention of similarly written codes (\cite{1997NewA....2..209A}, \cite{2011MNRAS.414.3458W}) gives a good balance of speed and accuracy.

The solution of the ODEs for ionization/heating is also parallelized. The computational volume is divided into sub-volumes and distributed between MPI processes, with multiple copies distributed to independent processes when there are a large number of sources. Each process then calls a CUDA kernel to assign these cells to the CUDA cores of each GPU. While the processes operate independently, MPI is used to consolidate data as necessary between ray tracing steps.

Once the calculation in each process is complete, the sub-volumes are recombined in the host process and redistributed between all processes. In some cases, the time required to distributed and recombine the grid data may be larger than the time required to actually perform the full ionization calculation; in these cases, we are also able to perform the ionization calculation locally in each process, eliminating the need to send any data between the processes during this step of the simulation.

\subsection{Heating and Cooling}
\begin{figure}
\includegraphics[width=8.6cm]{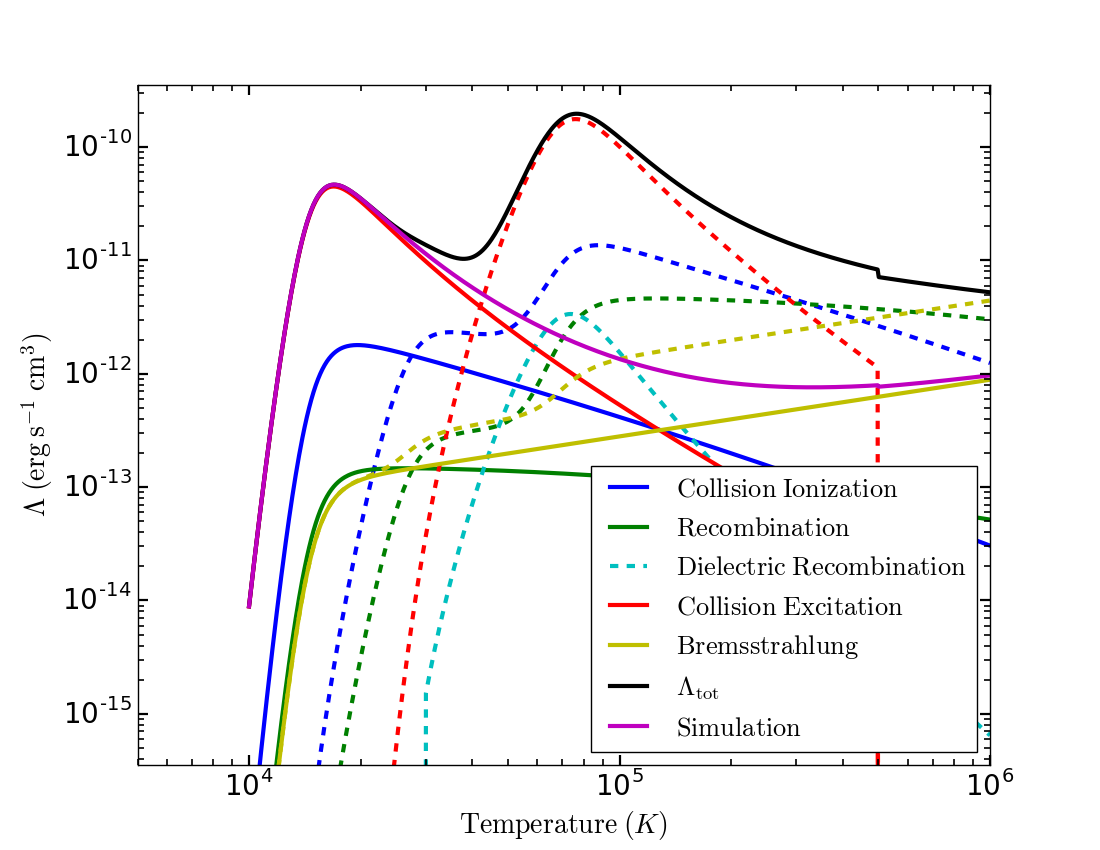}
\caption{Cooling function $\Lambda(T)$ (black solid line) as a function of temperature for a gas of primordial composition \citep{1997MNRAS.292...27H}. The solid lines represent the contribution to the cooling from different processes for hydrogen and the dashed lines for helium as shown in the legend.}
\label{fig:coolne}
\end{figure}

The heating of gas within each cell is calculated at the same time ionization rate is calculated. The energy per unit volume and per unit time added to each cell by a given ray is calculated based on the ionization rate:
\begin{align*}
\dot{E}_{\nu}=\sum_{\rm S} h(\nu-\nu_{\rm S})s_{\nu}n_{SI},
\end{align*}
where $\nu_{\rm S}$ is the ionization edge frequency of the species being tracked. We then use the available physical quantities (such as temperature, free electron density, ionized/neutral hydrogen/helium density) to calculate the cooling rate for a variety of physical processes, including collisional ionization, recombination, collisional excitation, bremsstrahlung, and adiabatic expansion due to the Hubble flow \citep{1997MNRAS.292...27H}. In Figure~\ref{fig:coolne}, we plot the cooling function we use in our simulations. In order to save memory, the simplest version of the code only tracks hydrogen and singly ionized helium fractions, meaning we assume a soft spectrum of radiation (stellar spectrum) in which \GIII number densities is negligible. In our initial simulations, all of the sources are initialized with spectra corresponding to stellar sources, so that temperatures above $2\times10^4$ K is rare, and we thus expect our cooling rates to be accurate. We initially omit tracking \GIII and primordial H$_2$ chemistry to save on computer memory; however, we have the ability to include other relevant chemistry, for instance when we include sources of X-rays and \pop3 stars that emit harder photons and $H_2$ dissociating radiation {(\rm cite)}.

\subsection{Radiative Transfer Time Step}

The previous section describes how we dynamically sub-cycle the time integration between radiative transfer calls; however, the choice of time step between calls of the radiative transfer routine is more difficult. The radiation field does not evolve between ray tracing calls, so that the velocity of the I-front is limited by the choice of this time step. This means that if the time step is chosen to be too large, the speed of the I-front is unphysically reduced. However, the I-front still approaches the correct asymptotic solution at large times regardless of choice of time step.

The ray tracing step is by far the most computationally expensive part of the simulation, so we wish to maximize the time step between ray tracing calls while keeping the required accuracy. In cases where the velocity of the ionization front is not important, we can opt to use a constant time step chosen based on physical considerations. In cases where higher accuracy is required, we have implemented two different schemes for determining the optimal time step adaptively:
\begin{enumerate}

\item {\bf Minimum neutral fraction change:} The simplest method for adaptively correcting the time step is regulate the maximum rate at which the neutral fraction in any given cell changes. For a single cell:
\begin{align*}
dt_n=\frac{\epsilon n_{\HI}}{|dn_{\HI}/dt|},
\end{align*}

where $\epsilon$ is the maximum fractional change in the neutral fraction. We calculate this quantity for every cell within the computational volume and calculate the minimum time step. We find that this method constrains the time step to be unnecessarily small in regions with small neutral fraction, where a large fractional change in neutral fraction results in a negligible change in the neutral density. We thus add the constraint to only consider cells with a relatively high optical depth:
\begin{align*}
\tau=n_{\HI}\sigma_id>0.5,
\end{align*}

where $d$ is the width of the cell and $\sigma_i$ is the cross section of the lowest frequency bin. This condition also limits us to cells near the ionization front.

\item {\bf Minimum intensity change:} Our second scheme for adaptively controlling the time step is to regulate the maximum rate at which the photon flux changes in a single cell. Similar to above, we define the time step for a given cell as:
\begin{align*}
dt_I&=\frac{\epsilon I_{\nu}}{|dI_{\nu}/dt|},
\end{align*}

Where $I_{\nu}=A\exp(-\tau_{\nu})$. We calculate this expression at runtime by using the following simplification:
\begin{align*}
\left|\frac{dI_{\nu}}{dt}\right|&=\left|\frac{d}{dt}A\exp(-\tau_{\nu})\right|\\
&=\left|A\exp(-\tau_{\nu})\frac{d\tau_{\nu}}{dt}\right|\\
&=\left|I_{\nu}\sigma_id\frac{dn_{\HI}}{dt}\right|,\\
dt_I&=\frac{\epsilon}{\sigma_id|dn_{\HI}/dt|}.
\end{align*}

We find that this method gives significantly larger time steps than the previous methods, with a minimal loss of accuracy when the ionization rate is very high. 

\end{enumerate}

We note here that for a given choice of time step, smaller \HII regions are underrepresented compared to larger \HII regions. We also note that the choice of time step makes much less of a difference in regions without radiation (as chemistry subcycles accurately solve the equations) and when \HII region overlap becomes prevalent (as most \HII regions are large and I-front velocities are low).

\subsection{Ionizing Background}

As the universe expands and the average electron fraction of the universe increases, the average density of neutral hydrogen decreases and the mean free path of photons in the IGM increases. These photons build up a ionizing background which becomes more dominant as more ionizing sources appear. The harder photons of the ionizing spectrum build up a background earlier than the softer photons due to their longer mean free path.
Individual halos never produce large enough regions where the gas is fully ionized to completely reionize the cosmic volume, so the derived average electron fraction underestimates the true electron fraction. We correct for this underestimation by calculating and including in the photon budget the ionizing background and its effect on the cosmic ionization history.
We quantify this effect solving the equation of radiation transfer in an homogeneous expanding universe as in \cite{Gnedin:2000, RicottiO:2004}, which we briefly summarize here. We begin with the number density of ionizing background photons $n_{\nu}$ at a redshift $z$ and evolve it to $z-\Delta z$. During each code timestep $\Delta z$, we add to the initial background at redshift $z$ (appropriately redshifted and absorbed by the neutral IGM) the photons produced by ionizing sources within our simulation between the redshifts of $z-\Delta z_0$ and $z-\Delta z$ including absorption and redshift effects (source term), where $z-\Delta z_0$ represents the redshift at which we begin adding the contribution of the sources to the background radiation. This parameter is used, as explained later, to avoid double counting the emission from low redshift (local) sources, that is already included in the radiation transfer calculation. Mathematically, we solve the equation:
\begin{align}
&n_{\nu}(z-\Delta z)=n_{\nu}(z)\exp\left[-\int_z^{z-\Delta z}dz'\alpha_{\nu'}(z')\right]\nonumber\\
&+\int_{z-\Delta z_0}^{z-\Delta z}dz' S_{\nu'}(z')\exp\left[-\int_{z'}^{z-\Delta z}dz''\alpha_{\nu''}(z'')\right],\label{eq:background}
\end{align}
where $\nu'=\nu (1+z')/(1+z)$ we have defined a dimensionless absorption coefficient and source function:
\begin{align}
&\alpha_{\nu}=\frac{(1+z)^2}{H(z)}c{\overline n}_{H}\sigma_{\nu}(\HI)(1-x_e(z)),\\
&S_{\nu}=\frac{{\dot n}_{\rm ion}\langle h\nu\rangle g_\nu/h\nu}{(1+z)H(z)},
\end {align}
where the sources spectra are normalized as $\int_{\nu_0}^\infty g_\nu d\nu=1$, with $h\nu_0=13.6$~eV and $\langle h\nu\rangle^{-1} \equiv \int_{\nu_0}^\infty (g_\nu/h\nu) d\nu$. We use this equation to calculate two quantities which we track throughout the simulation:

\begin{enumerate}
\item The overall background $n_{\nu}^{\rm all}(z-dz)$ is calculated using $\Delta z_0=0$, so that we include background and local radiation. This quantity is used as $n_{\nu}(z)$ in Equation \ref{eq:background} for the next time slice.
\item The mean local radiation emission $n_{\nu}^{\rm loc}(z-dz)$ is calculated assuming $\Delta z_0=H_0R_0/c$, where $R_0$ represents the size of the simulation box. 
By combining the local radiation background (from stars inside the box but within $\Delta z$) with the overall background $n_{\nu}^{\rm all}$ from previous slices, we find the background radiation from all stars at higher redshifts in the simulation without double counting the contribution of sources inside the box.
\end{enumerate}

\subsection{MPI/CUDA Parallelization}\label{sec:parallel}

The $N^3$ uniform grid of our full simulation represents a simple first application of our ray tracing algorithm. In the first version of ARC (v1) every GPU requires the full grid data to perform the ray tracing calculations, which limits the possible size of the simulation grid. In the current version (v2), if the grid required for the simulation is larger than can be stored on a single GPU, the code allow the code to break the volume into smaller regions. When rays reach the edge of a sub-volume, they are sent to the adjacent sub-volume into which they move, much in the same way as AMR methods perform the calculation. This means our method is applicable in AMR settings, though our current iteration of the code is limited to Cartesian grids which may be subdivided. Regardless of these considerations, GPUs excel at performing these ray tracing calculations as long as the number of rays within a single computational volume is large and the CUDA core occupancy is near optimal ($>10^4$).


\section{Radiative Transfer Tests}\label{sec:tests}

Consistency tests are paramount in demonstrating that a new radiative transfer code produces results that are consistent with well established methods or scenarios that have exact analytical solutions. In this section we present the results produced by our code when running the tests presented in \citep{2006MNRAS.371.1057I} (RT06). These tests include 0) Tracking chemistry in a single cell 1) expansion of an isothermal \HII region in pure hydrogen gas 2) expansion of an \HII region in a pure hydrogen gas with evolving temperature 3) I-front trapping and the formation of a shadow 4) multiple sources in a cosmic density field. We show that our code performs well when compared to CPU codes that use similar ray-tracing methods. The various codes used for comparison in RT06 show a good deal of variability for several of the tests. The method implemented in our code most closely resembles $C^2$-Ray \citep{2006NewA...11..374M} and MORAY \citep{2011MNRAS.414.3458W}. Where possible we compare our results to the publicly available RT06 results for $C^2$-Ray.
Since at the moment our ray tracing algorithm is used in a post-processing setting, it neglects hydrodynamic evolution. We therefore limit ourselves to tests which do not require accurate tracking of the hydrodynamic properties of the gas.

\subsection{Test 0 - Chemistry in a Single Cell}

The test presented in RT06 applies a constant radiation field of plane-parallel radiation to a single cell of the simulation. Our code is limited to tracking point sources of radiation, so we place a single source of photons at one side of a 6.6 kpc cube with $128^3$ cells. The source is given a luminosity $5.2\times10^{57}\;{\rm photons}\;{\rm s}^{-1}$ so that the flux at the cell is $10^{12}\;{\rm photons}\;{\rm s}^{-1}\;{\rm cm}^{-2}$. The density of the cell is $n=1\;{\rm cm}^{-3}$ and it is initially neutral at temperature $T=100$ K. The radiation is approximated to be a blackbody of temperature $10^5$ K; following \cite{2011MNRAS.414.3458W}, we use four frequency bins with central energies $E_i=(16.74, 24.65, 34.49, 52.06)$ and relative luminosities $L_i/L=(0.277, 0.335, 0.2, 0.188)$. The radiation is applied for 0.5 Myr, after which the radiation is turned off and the cell is tracked for 5 Myr.

\begin{figure}[t]
\includegraphics[width=8.6cm]{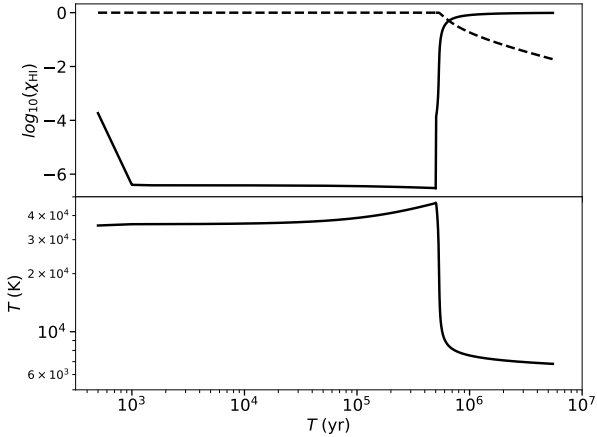}
\caption{Test 0 - Chemistry in a single cell. In this test, a single cell is subject to intense radiation for $5\times10^5$ yr, at which point the radiation stops. ({\it Top.}) Plot of the neutral fraction (solid line) and ionized fraction (dashed line) as a function of time. ({\it Bottom.}) Plot of the temperature as a function of time. These results agree well with the results presented in RT06, except at the first time step, which is expected given our use of a fixed time step.}
\label{fig:T0_0}
\end{figure}

In figure \ref{fig:T0_0} we plot the neutral fraction and temperature of the cell as a function of time. We find that these results agree with those presented in RT06. The only discrepancy between our results and those presented in RT06 is in the electron fraction for the first time step. This is a result of the optically thin approximation for sub-cycles. The first ray tracing calculation occurs at $x_{\HI}=1$, at which point the cell absorbs more soft photons, so that the chemistry calculator doesn't reach the correct equilibrium point. After this step, however, the optically thin approximation is satisfied and the solution agrees from the second step on. 

\subsection{Test 1 - Expansion of Isothermal \HII region in Pure Hydrogen}
\begin{figure*}[tbh]
\includegraphics[width=8.0cm]{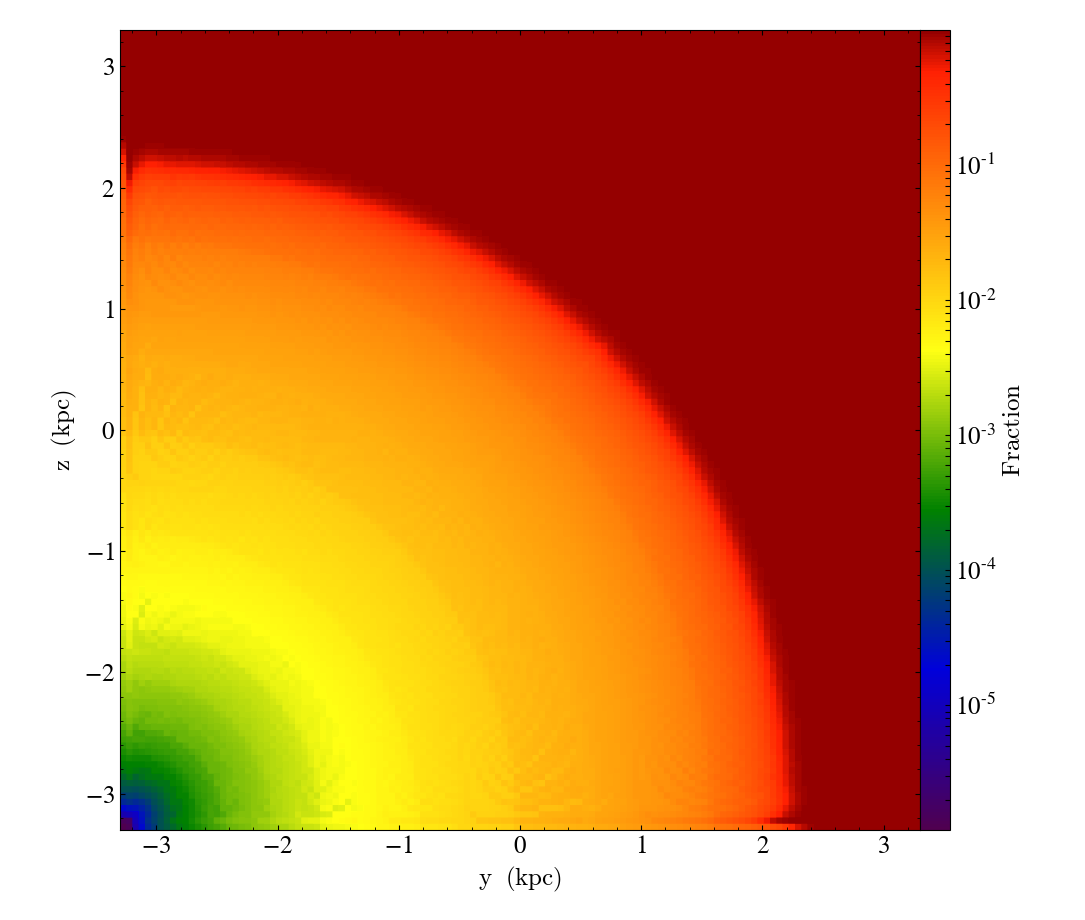}~~
\includegraphics[width=9.2cm]{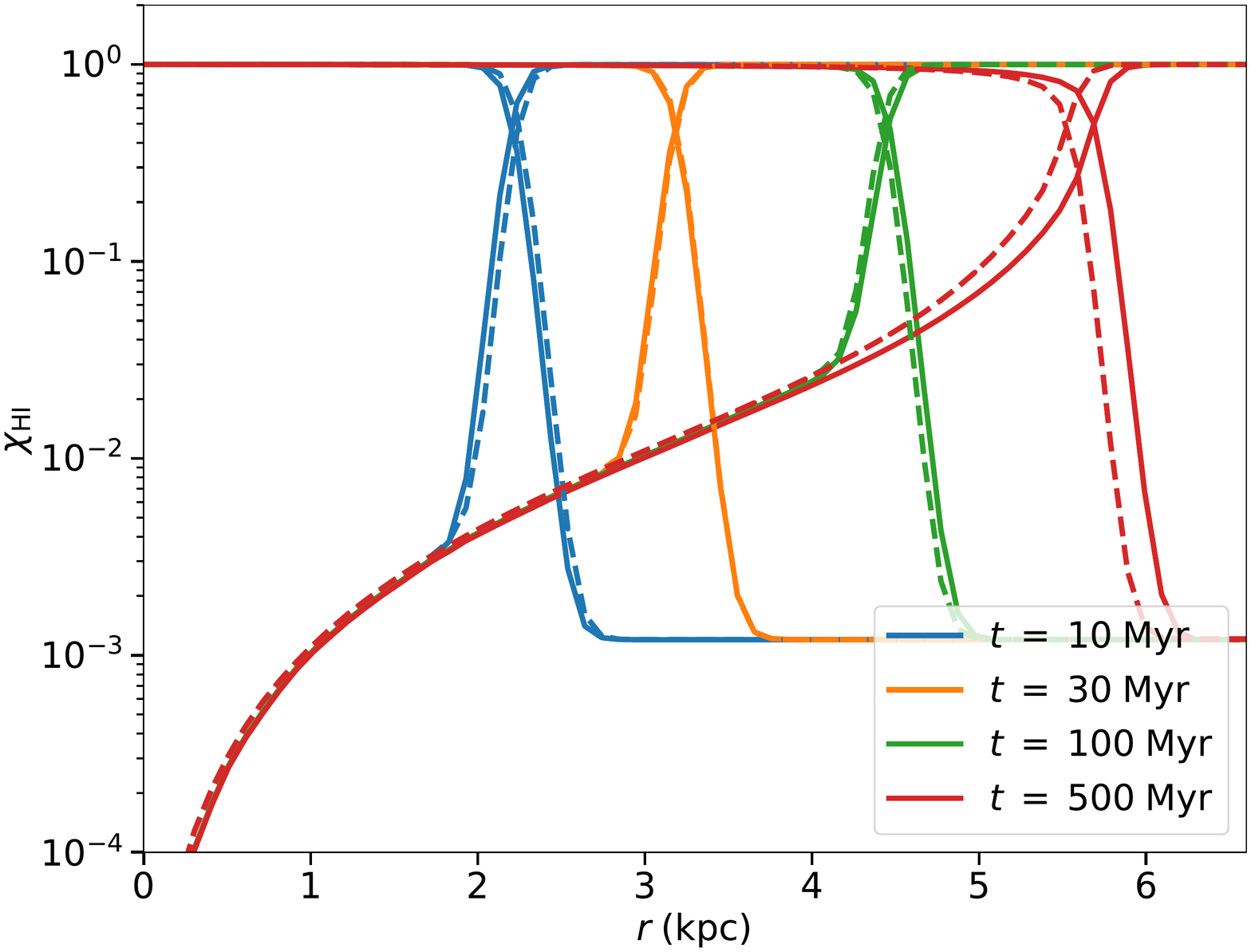}
\includegraphics[width=8.9cm]{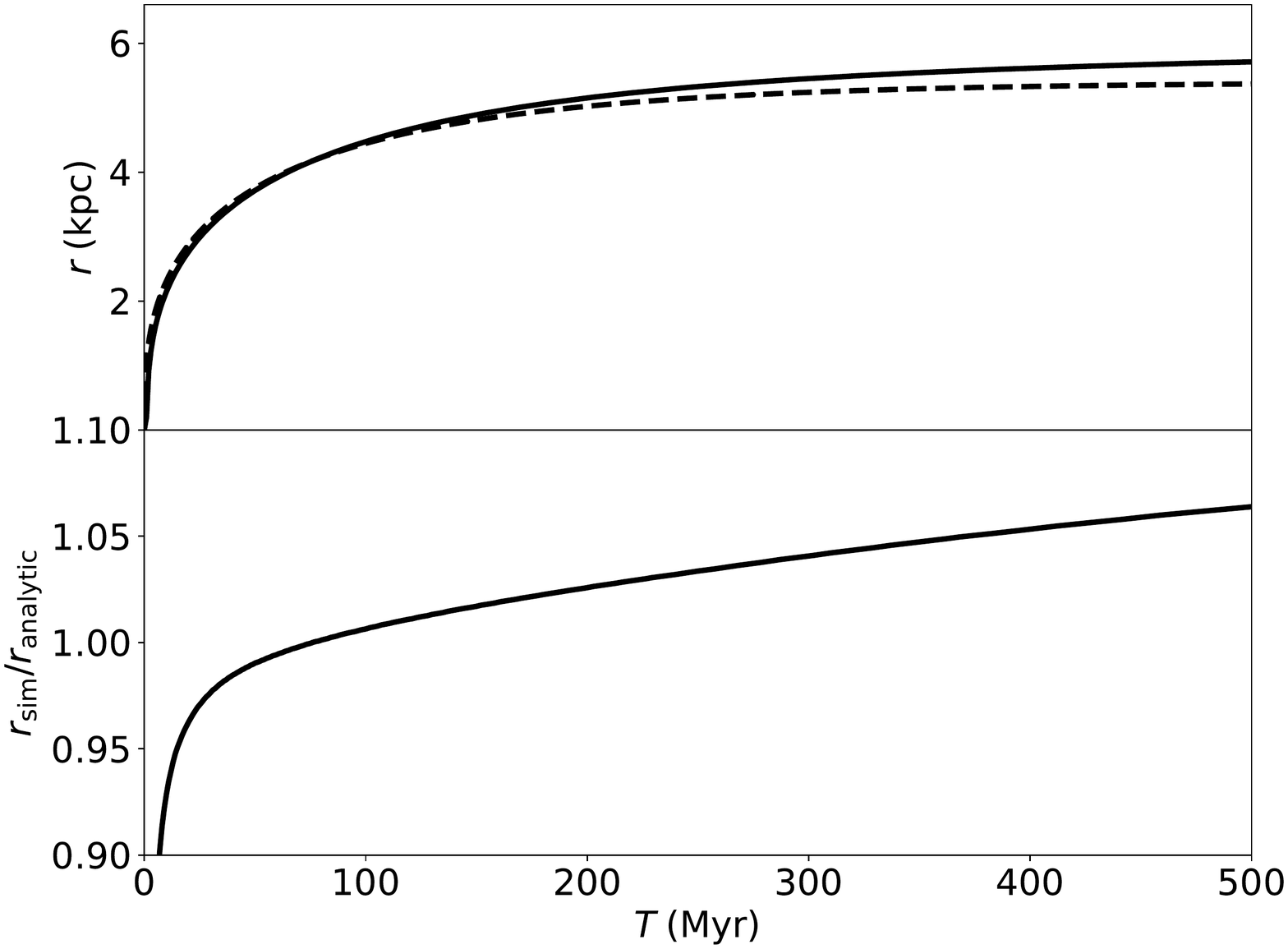}~
\includegraphics[width=8.9cm]{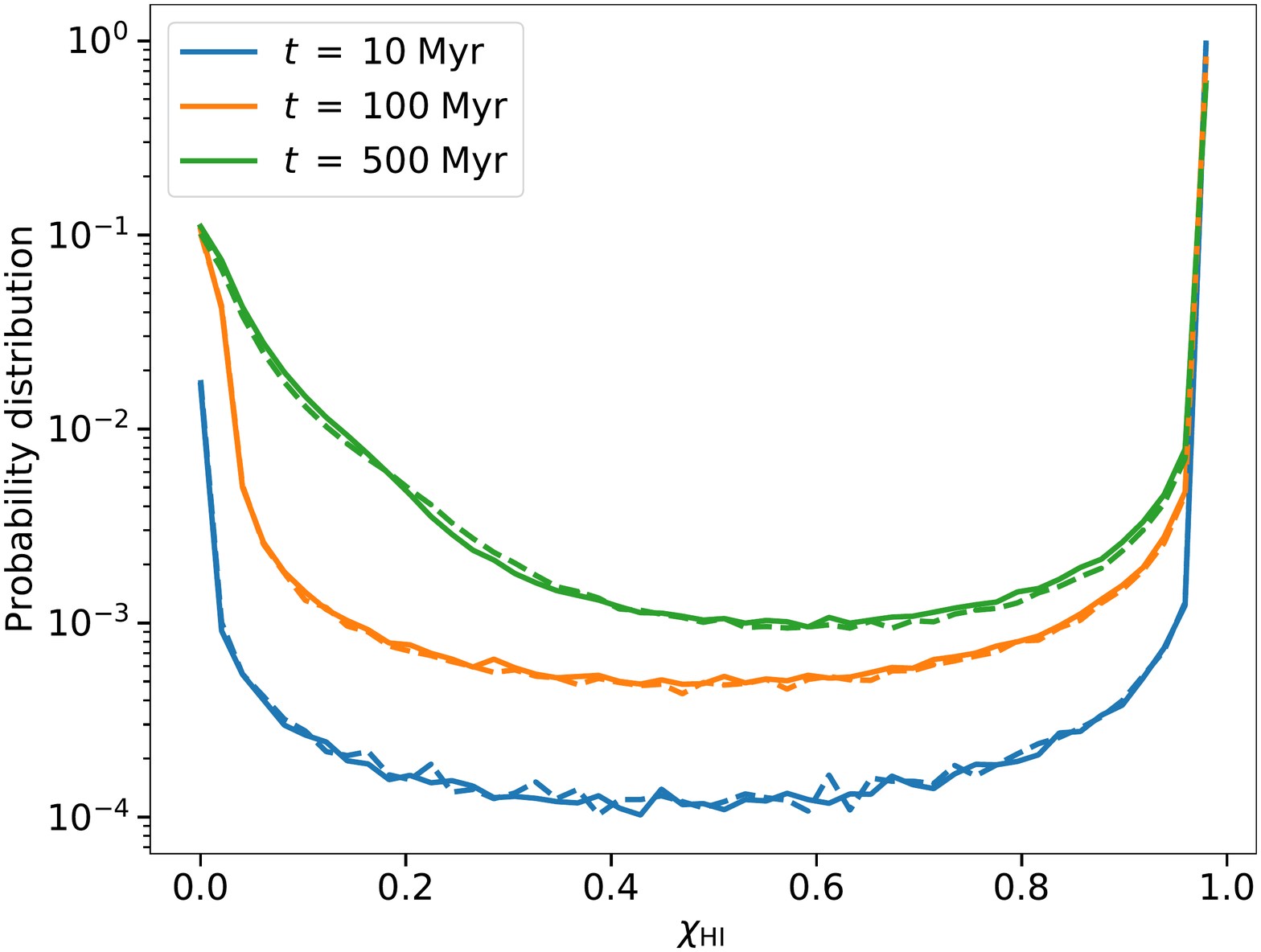}
\caption{Test 1 - Expansion of an isothermal \HII region in a gas of pure hydrogen. ({\it Top-left.}) Slice plots of the neutral fraction at 500 Myr into the simulation. The size and shape of the \HII region are in good agreement with the results in RT06. ({\it Top-right.}) Radially averaged neutral fraction as a function of distance from the source, in kpc. The dashed lines are the same plots reproduced from RT06 using $C^2$-Ray. ({\it Bottom-left.}) Ionization front plotted along with the analytical model (top) and the ratio of these models (bottom). We note that the simulation deviates from the analytic solution at the beginning and end of the simulation. This is a result of the soft 13.6 eV spectrum being the worst case scenario for our optically thin approximation for non-equilibrium chemistry calculation. ({\it Bottom-right.}) Histogram of neutral fractions at $t=10,\;100,\;500$  Myr. The dashed lines are the same plots reproduced from RT06 using $C^2$-Ray. We see that our model matches $C^2$-Ray very well.}
\label{fig:T1_0}
\end{figure*}

The most fundamental test for a radiative transfer code is the simulation of an expanding \HII region around a constant luminosity source in a constant density medium of pure hydrogen 
Assuming the medium is initially neutral and that the \HII region has a sharp boundary, we have the well known analytic solution:
\begin{align}\label{eq:strom}
r_I(t)&=r_S\;\left(1-\exp\left(\frac{t}{t_{\rm rec}}\right)\right)^{1/3},\\
v_I(t)&=\left(\frac{r_S}{3t_{\rm rec}}\right)\frac{\exp(t/t_{\rm rec})}{(1-\exp(t/t_{\rm rec}))^{2/3}},
\end{align}

where:
\begin{align*}
r_S&=\left(\frac{3\dot{N}}{4\pi\alpha_B(T)n_H^2}\right)^{1/3},\\
t_{\rm rec}&=[\alpha_B(T)n_H]^{-1}.
\end{align*}

The test domain is a 6.6 kpc cube with a $128^3$ element cubic grid. The gas has a fixed density $n_H=1.0\times10^{-3}$ and temperature $T=10^4$ K and an initial equilibrium ionization fraction of $1.2\times10^{-3}$. We place a single ionizing source in the corner of the simulation box, and assume a monochromatic spectrum with energy $h\nu = 13.6$ eV and luminosity of $S_0=5\times10^{48}\;{\rm photons}\;{\rm s}^{-1}$. The chosen physical parameters give a Str{\"o}mgren radius $R_S=5.4$ kpc and a recombination time $t_{\rm rec}=122.4$ Myr. We track the evolution of this simulation for 500 Myr, or roughly four recombination times.
\begin{figure*}[tbh]
\centering
\includegraphics[width=8.7cm]{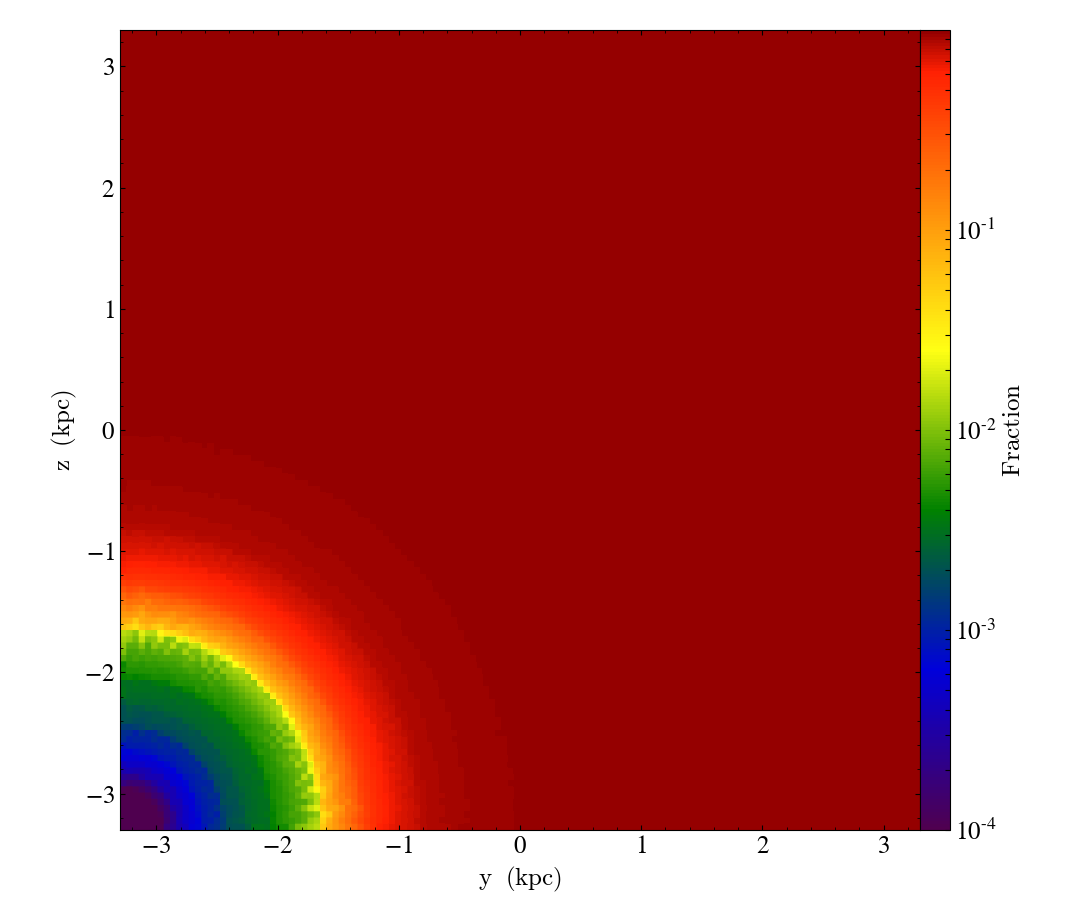}
\includegraphics[width=8.7cm]{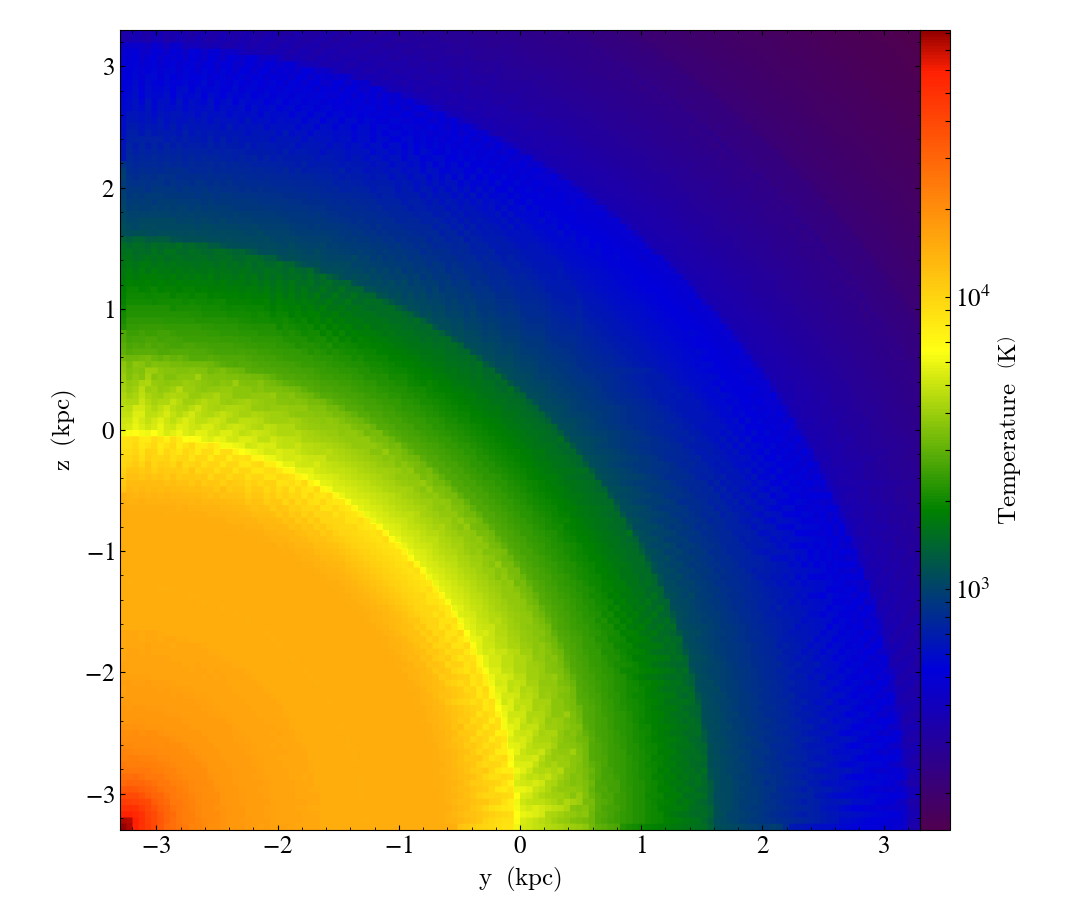}\\
\includegraphics[width=8.7cm]{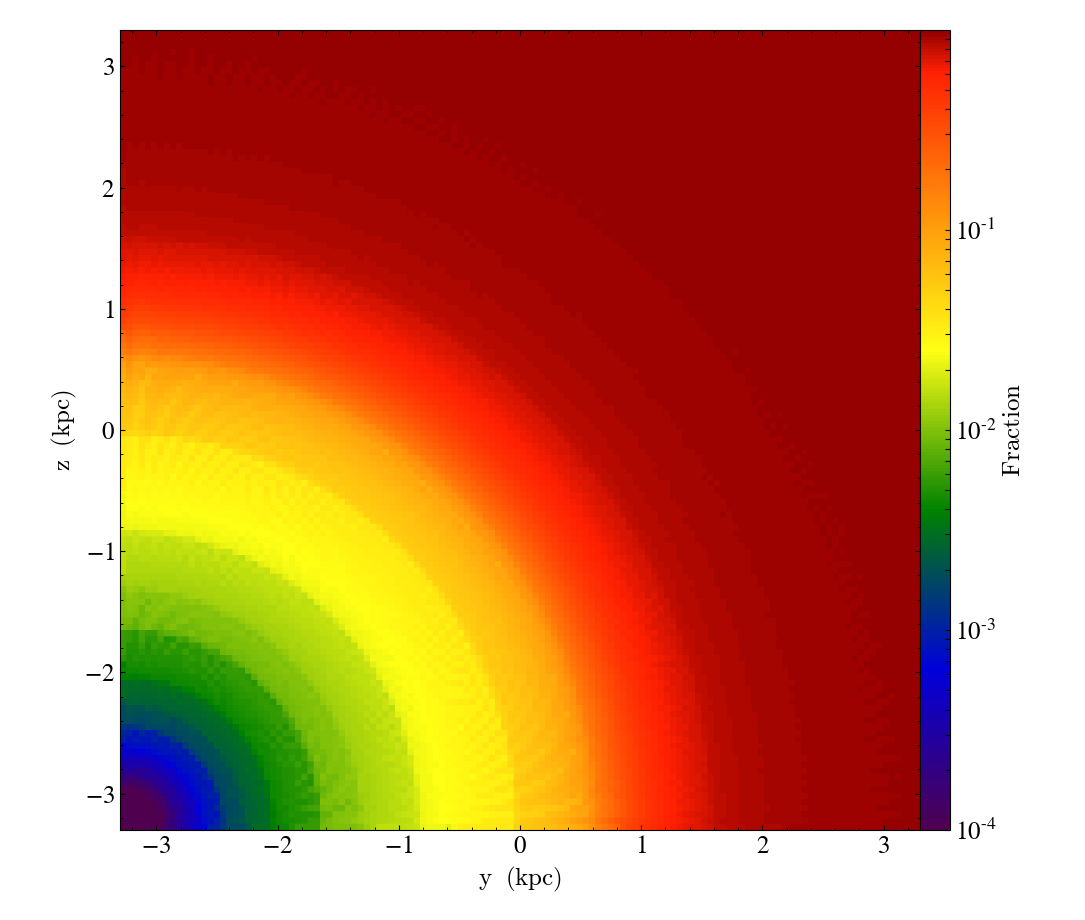}
\includegraphics[width=8.7cm]{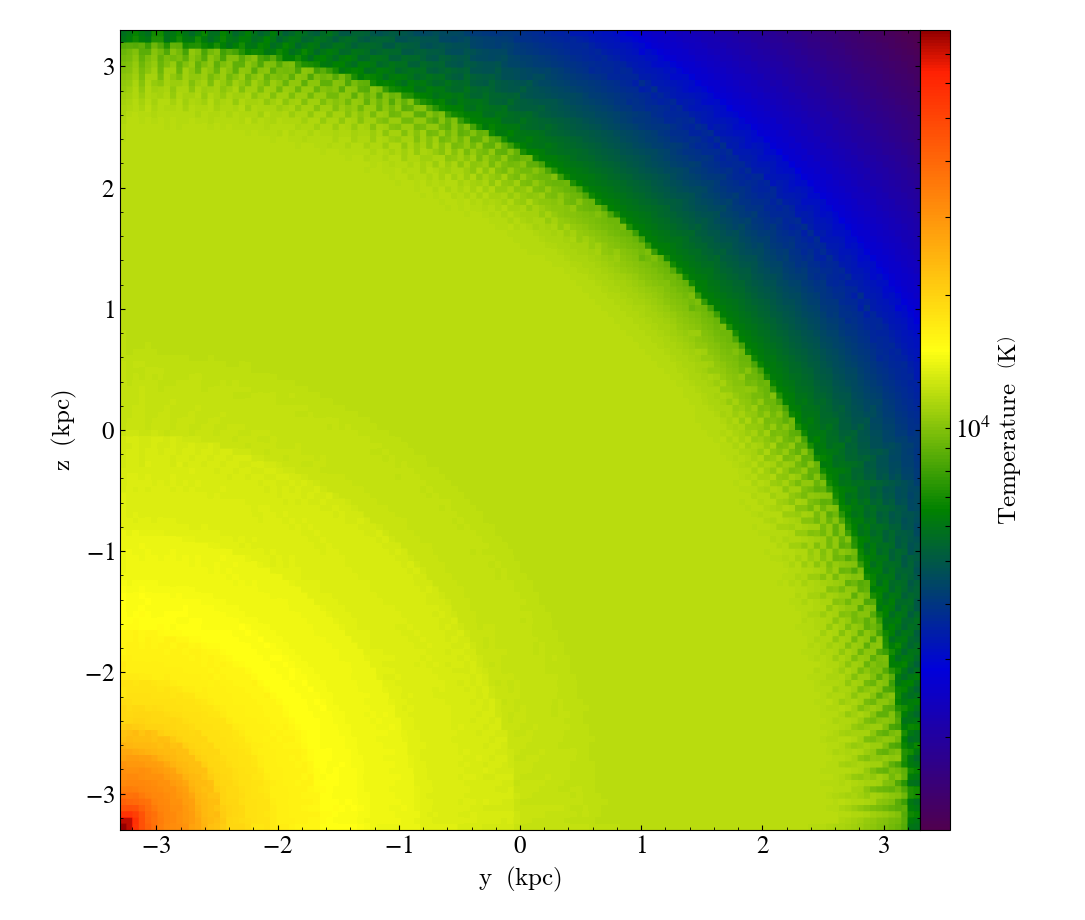}
\caption{Test 2 - Expansion of \HII region in a gas of pure hydrogen with evolving temperature. ({\it Top and bottom left.}) Slice plots of the neutral fraction at 10~Myr and 500~Myr into the simulation, respectively. ({\it Top and bottom right.}) Slice plots of the temperature at 10~Myr and 500~Myr into the simulation. The adopted $10^5$~K blackbody spectrum introduces harder photons which heat the gas to a much larger radius than the size of the Str{\"o}mgren sphere, in agreement with the results presented in RT06.}
\label{fig:T2_0}
\end{figure*}

In Figure~\ref{fig:T1_0} we plot a slice through the computational volume at t=500 Myr (top left). This panel shows that the code produces a spherically growing region of ionization, as expected. In the top-right panel of Figure~\ref{fig:T1_0} we plot the radially averaged profile of the \HI fraction as a function of radius at four times throughout the simulation. The shape of this profile agrees well with the results of RT06. In the bottom-left panel of Figure~\ref{fig:T1_0} we plot the comparison of our simulation with the model in Equation~\ref{eq:strom}. We see that our model agrees with the analytic fit within 5\% for the majority of the simulation. The underestimation of the radius at early times is a result of using a fixed time step for these tests; with an adaptive time step, the agreement is much better (see Figure \ref{fig:rad_correction}). Finally, in the bottom-right panel we show the histogram of \HI fraction for three times in the simulation (solid lines). The dashed lines represent the same plots produced using the $C^2$-Ray data from RT06. We see that our code produces the expected results for this test.


\subsection{Test 2 - Expansion of \HII Region in Pure Hydrogen with Evolving Temperature}
\begin{figure*}
\includegraphics[width=8.9cm]{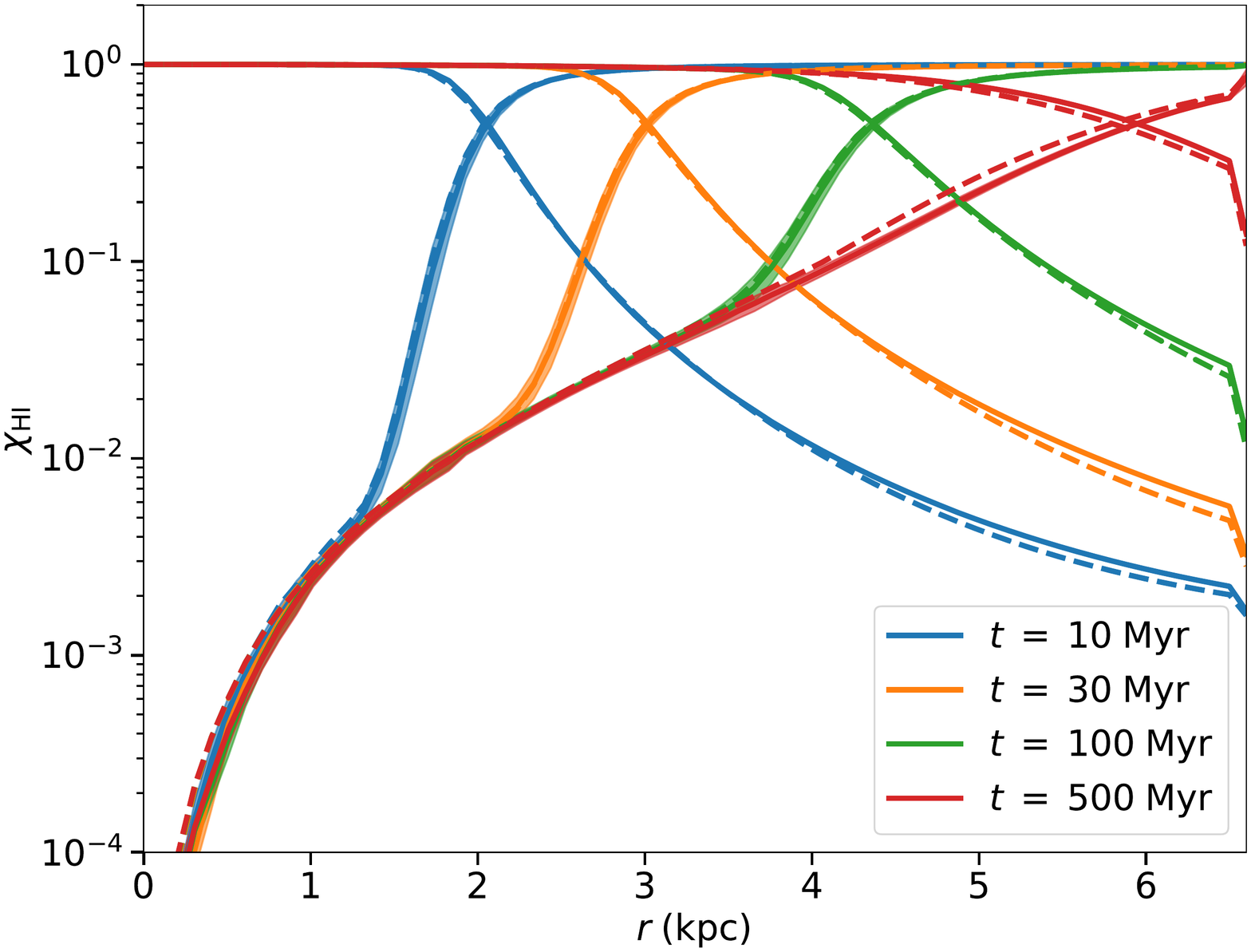}
\includegraphics[width=8.9cm]{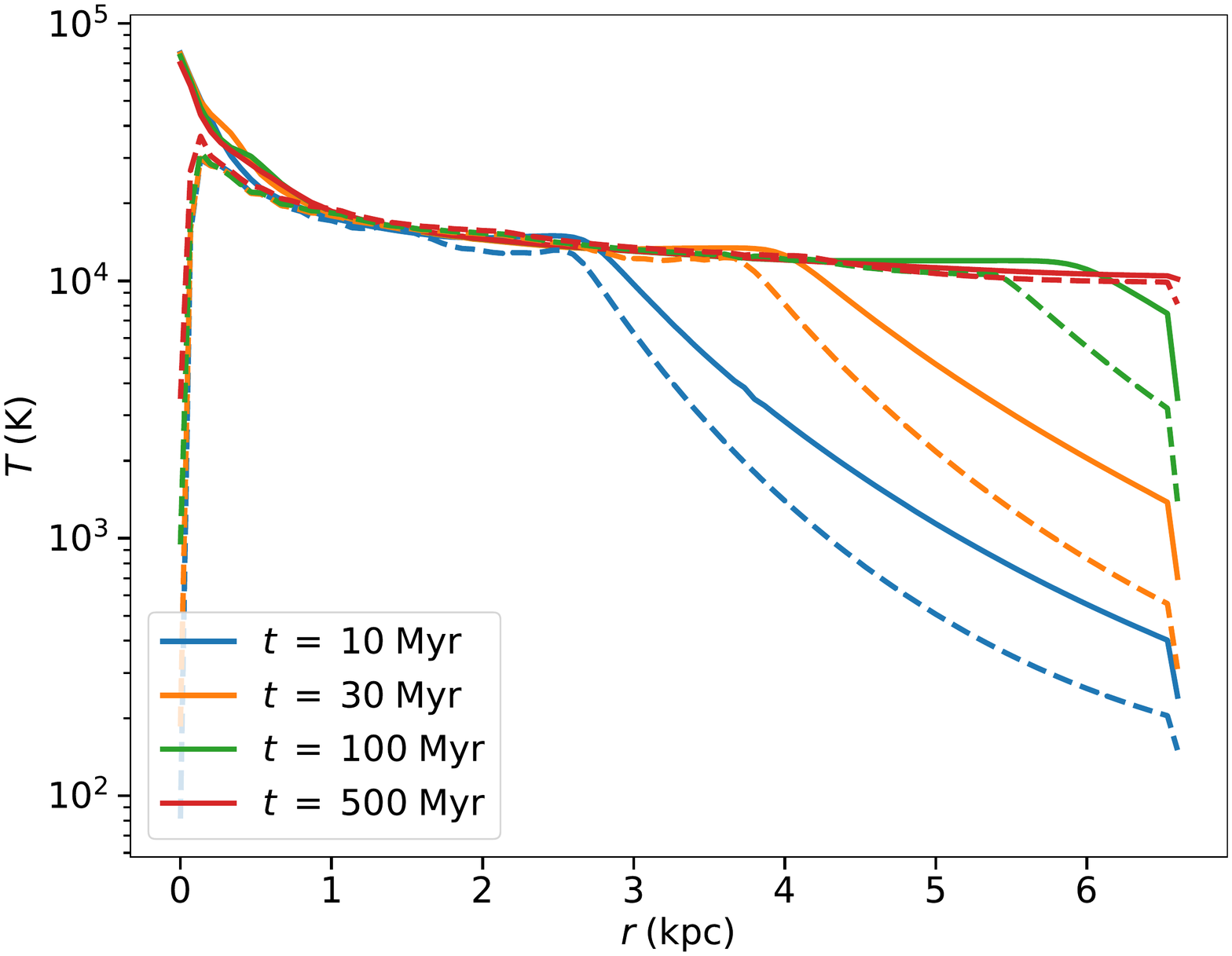}
\includegraphics[width=8.9cm]{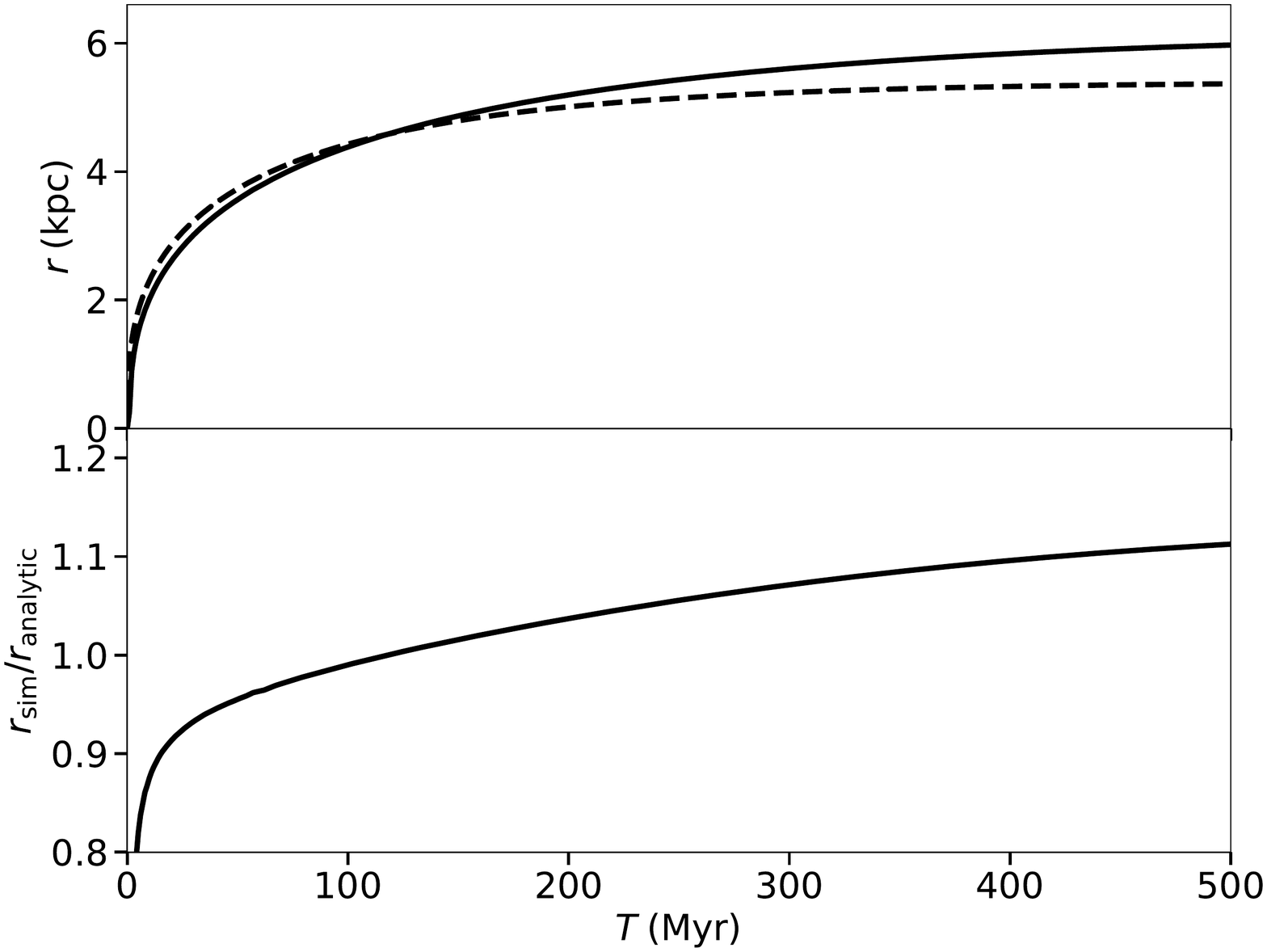}
\includegraphics[width=8.9cm]{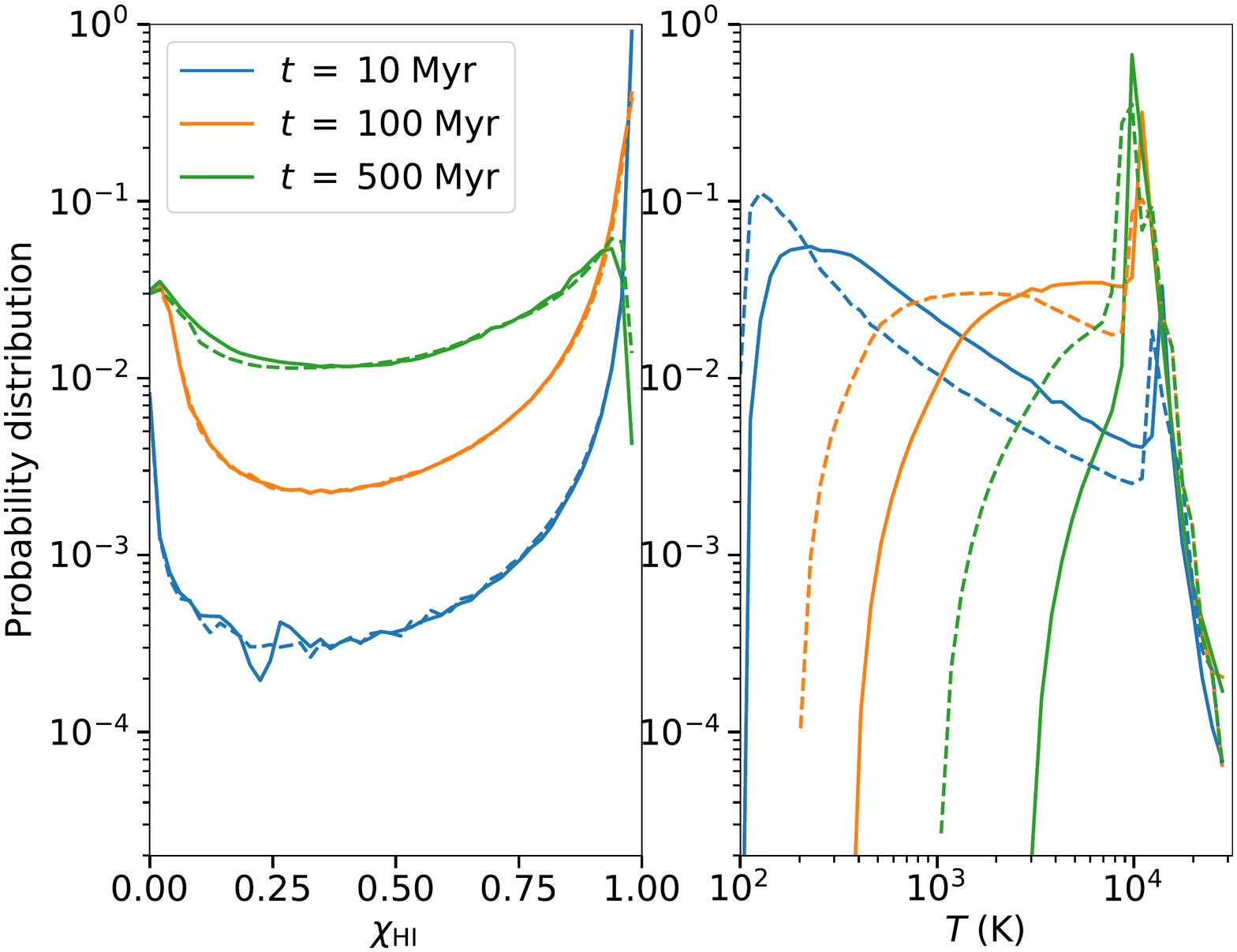}
\caption{Test 2 - Expansion of \HII region in a gas of pure hydrogen with evolving temperature. Radially averaged neutral fraction ({\it top-left}) and temperature ({\it top-right}) as a function of distance from the source, in kpc. The shaded thickness of the solid neutral lines represents the variance of the radial average. The dashed lines in both plots represent the same plots reproduced from RT06 using $C^2$-Ray. The slight disagreements are a result of a different assumption on the spectrum of the source (monocromatic vs blackbody spectrum) between our code and the $C^2$-Ray run in RT06.
({\it Bottom-left.}) Ionization front plotted along with the analytical isothermal-model (top) and the ratio of these models (bottom). This model deviates more than in test 1, a result of the non-isothermal nature of the simulation. ({\it Bottom-right.}) Histograms of the neutral fraction and temperature within the simulation, respectively. The dashed lines represent the same results calculated from $C^2$-Ray data from RT06. We find that our code produces slightly higher temperatures than $C^2$-Ray, a result of the difference assumption on the source spectrum between our code and the $C^2$-Ray run used in RT06.
}
\label{fig:T2_1}
\end{figure*}

The second test is similar to the first test, but with the reintroduction of heating and cooling processes to the simulation. We give the source a blackbody spectrum with $T=10^5$ K, using the same spectrum and luminosity as Test 0. The gas is given an initial temperature $T=100$ K.

In the top and middle left panels of Figure~\ref{fig:T2_0} show a slice through the neutral fraction at $t=10$ Myr and $t=100$ Myr, respectively. We see that the boundaries of the \HII region in these plots is less sharp than those of Fig.~\ref{fig:T1_0}, as anticipated with the harder radiation present in the $T=10^5$ K black body spectrum. We plot $t=100$~Myr instead of $t=500$~Myr for this test because the edge of the \HII region and heating reaches the boundary by then, meaning the plots show less relevant information. In the top and middle right panels of the same figure, we plot a slice through the temperature at the same times to show how the radiation is able to heat at a larger radius than the ionization front. In Figure~\ref{fig:T2_1} we show the respective radial profiles of the neutral fraction (top-left) and the temperature (top-right). We plot our results (solid lines) against the RT06 results for $C^2$-Ray (dashed lines) for comparison. In contrast to Figure~\ref{fig:T1_0}, we see that the neutral fraction increases more gradually, in good agreement with the results of $C^2$-Ray. We note that our model heats the gas outside of the Str{\"o}mgren sphere more than the $C^2$-Ray results from RT06. This is a result of the difference between the blackbody spectrum from \cite{2011MNRAS.414.3458W} and the one used in RT06.

In the bottom-left panel of Figure~\ref{fig:T2_1} we show the growth of the radius of the Str{\"o}mgren sphere as a function of time. We see that the radius of the ray tracing model begins lagging behind the analytic model while the gas is being heated, and later moves beyond the analytic model once the gas is heated beyond $10^4$ K. These results are in good agreement with the models presented in RT06. Finally, the bottom-right panel of Figure~\ref{fig:T2_1} shows the histogram of neutral fractions (left plot) and temperatures (right plot) at $t=10$, 100, and 500 Myr (solid lines) against the $C^2$-Ray results (dashed lines). Again, we see the neutral fractions are in excellent agreement, while our model heats the lower temperature gas slightly more, due to the adoption of a harder spectrum that was chosen to match the one adopted in \cite{2011MNRAS.414.3458W}.

\subsection{Test 3 - I-front Trapping and Formation of a Shadow}
\begin{figure*}[tbhp]
\centering
\includegraphics[width=8.5cm]{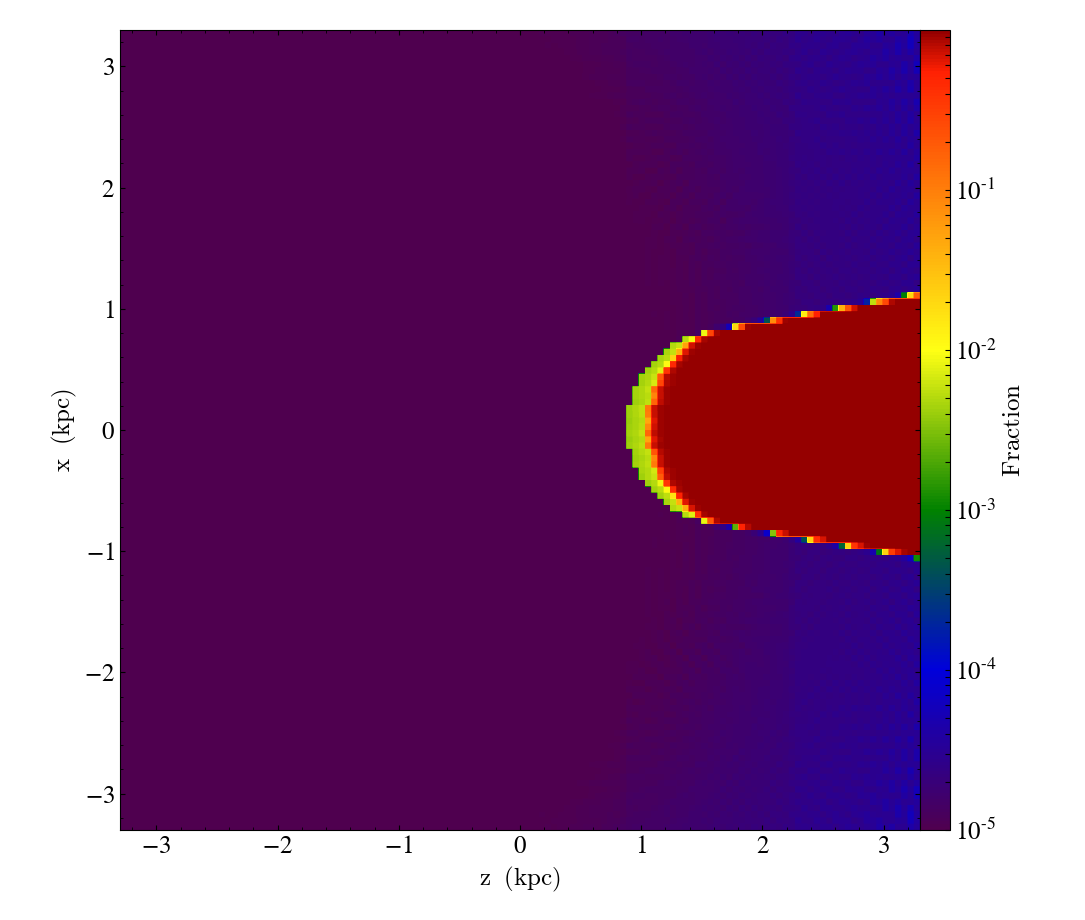}
\includegraphics[width=8.5cm]{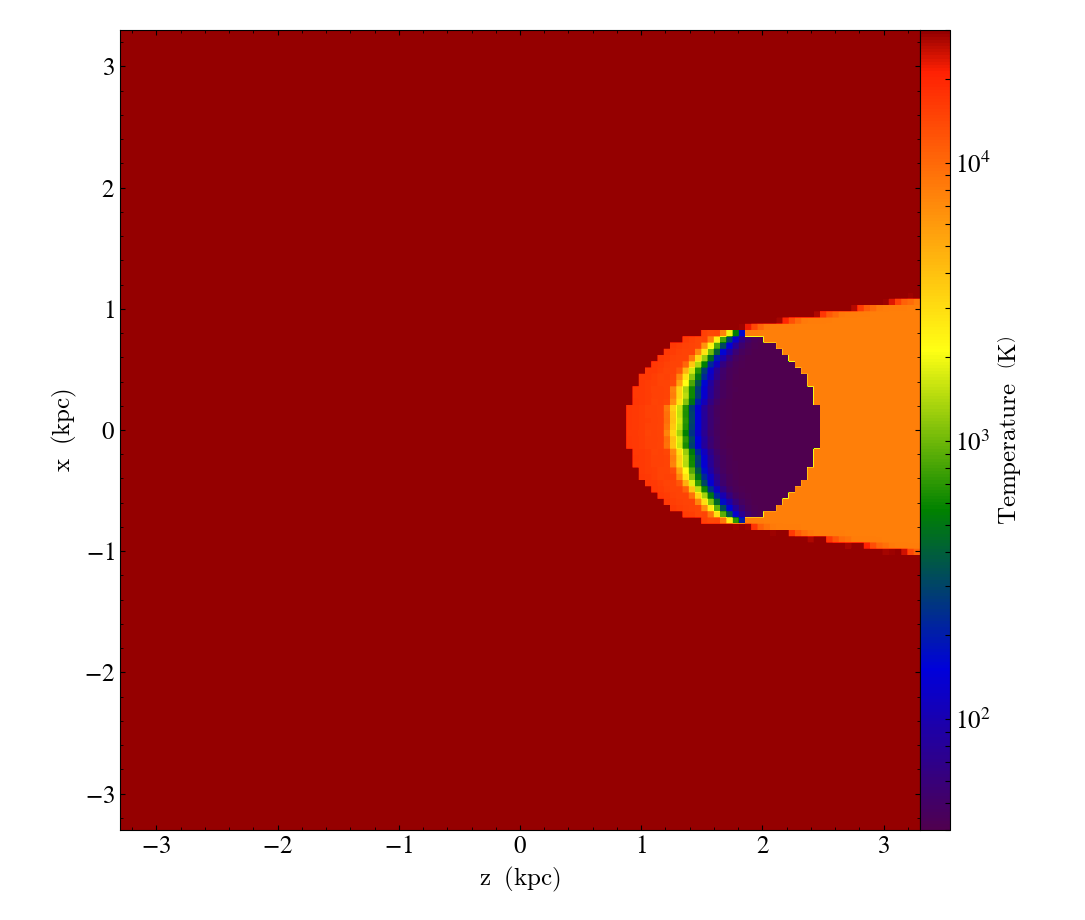}\\
\includegraphics[width=8.5cm]{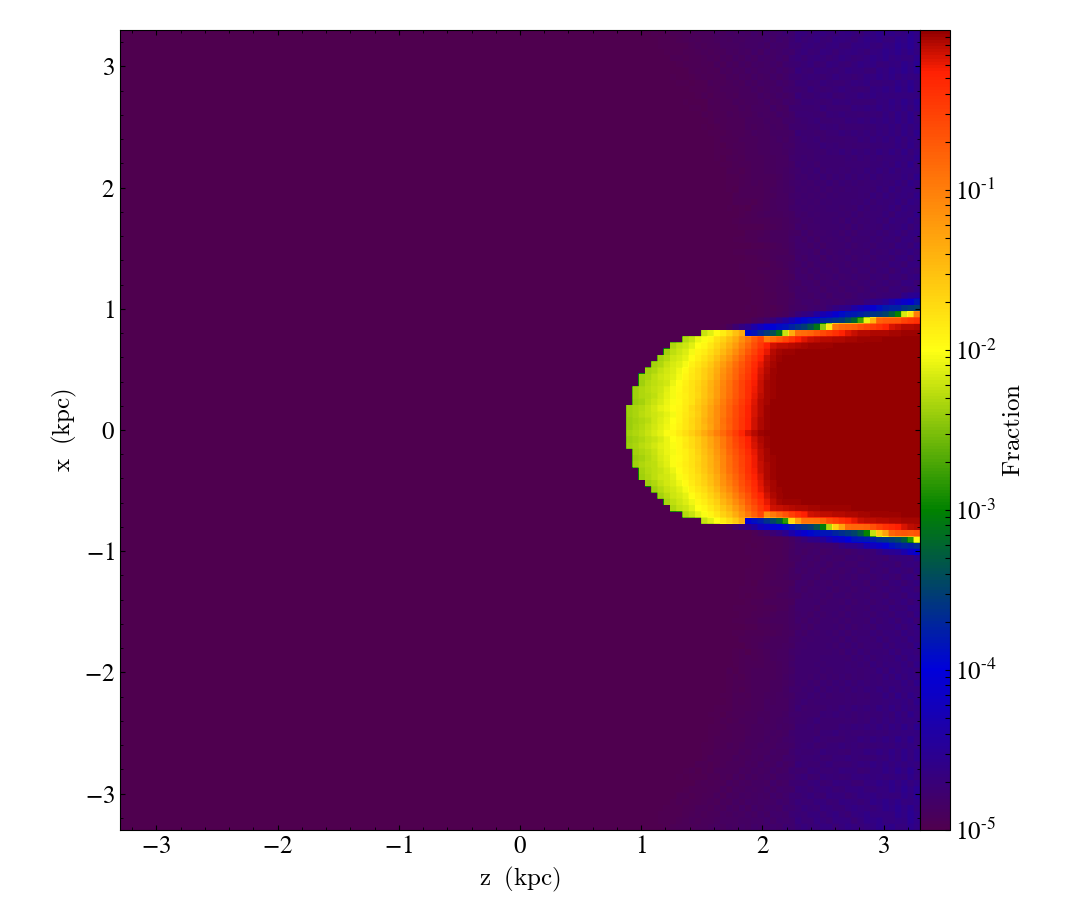}
\includegraphics[width=8.5cm]{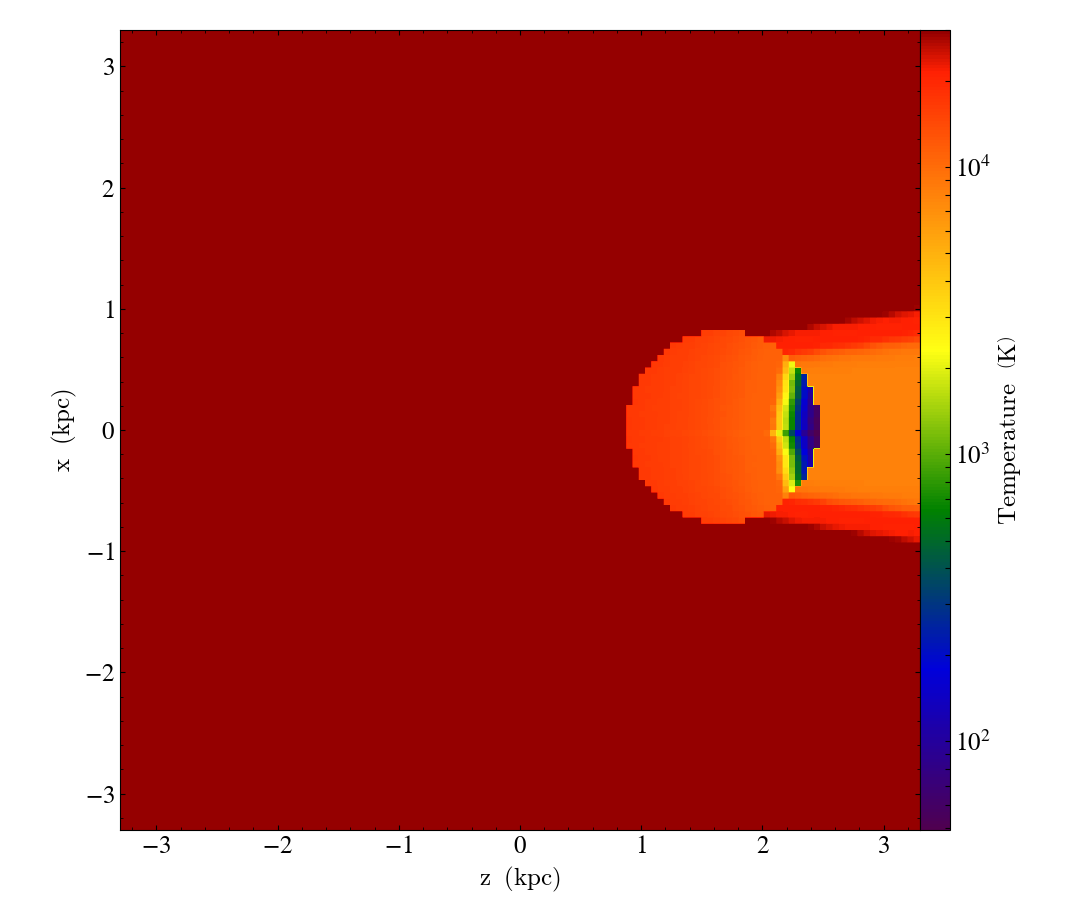}\\
\includegraphics[width=8.5cm]{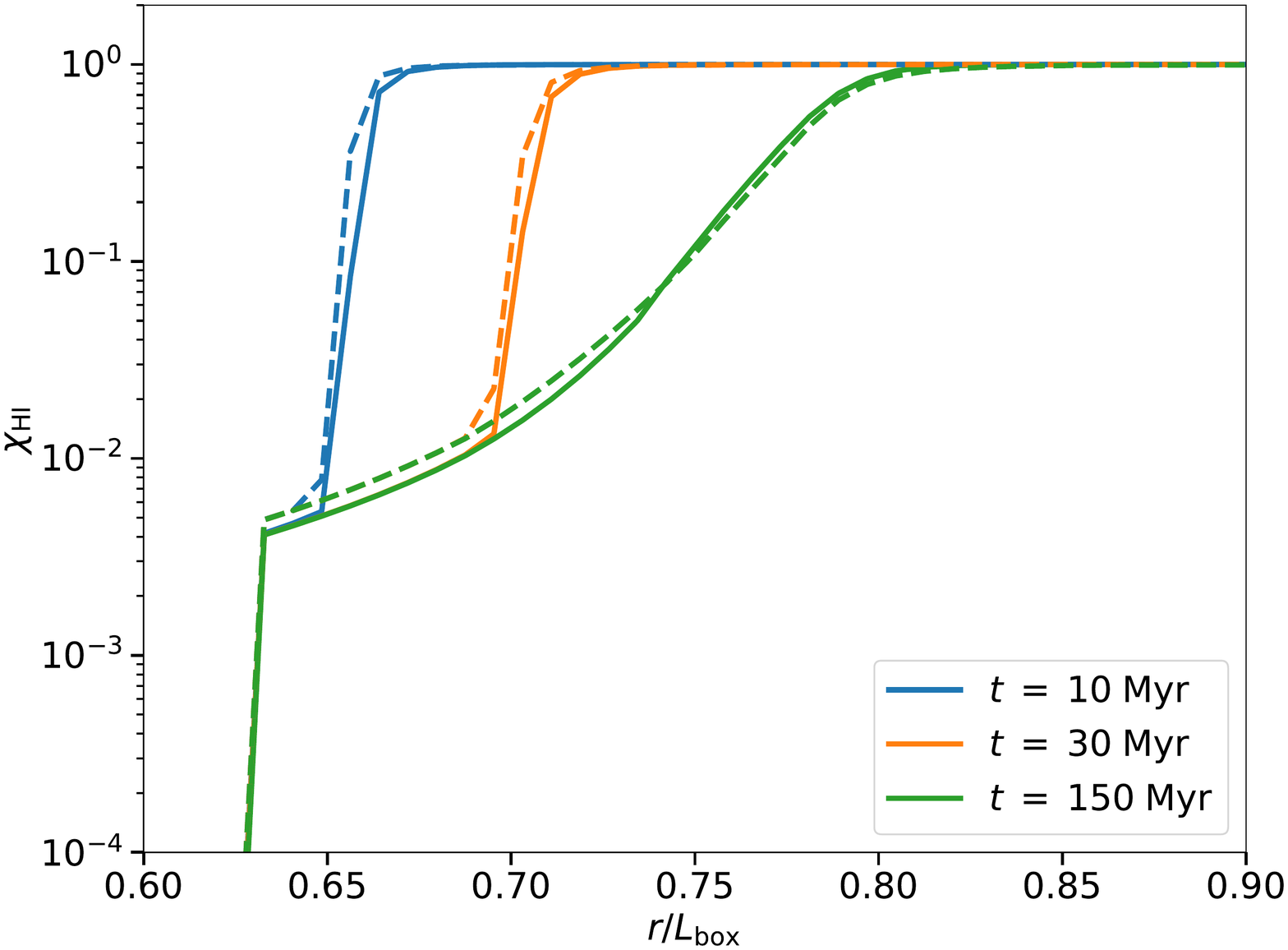}
\includegraphics[width=8.5cm]{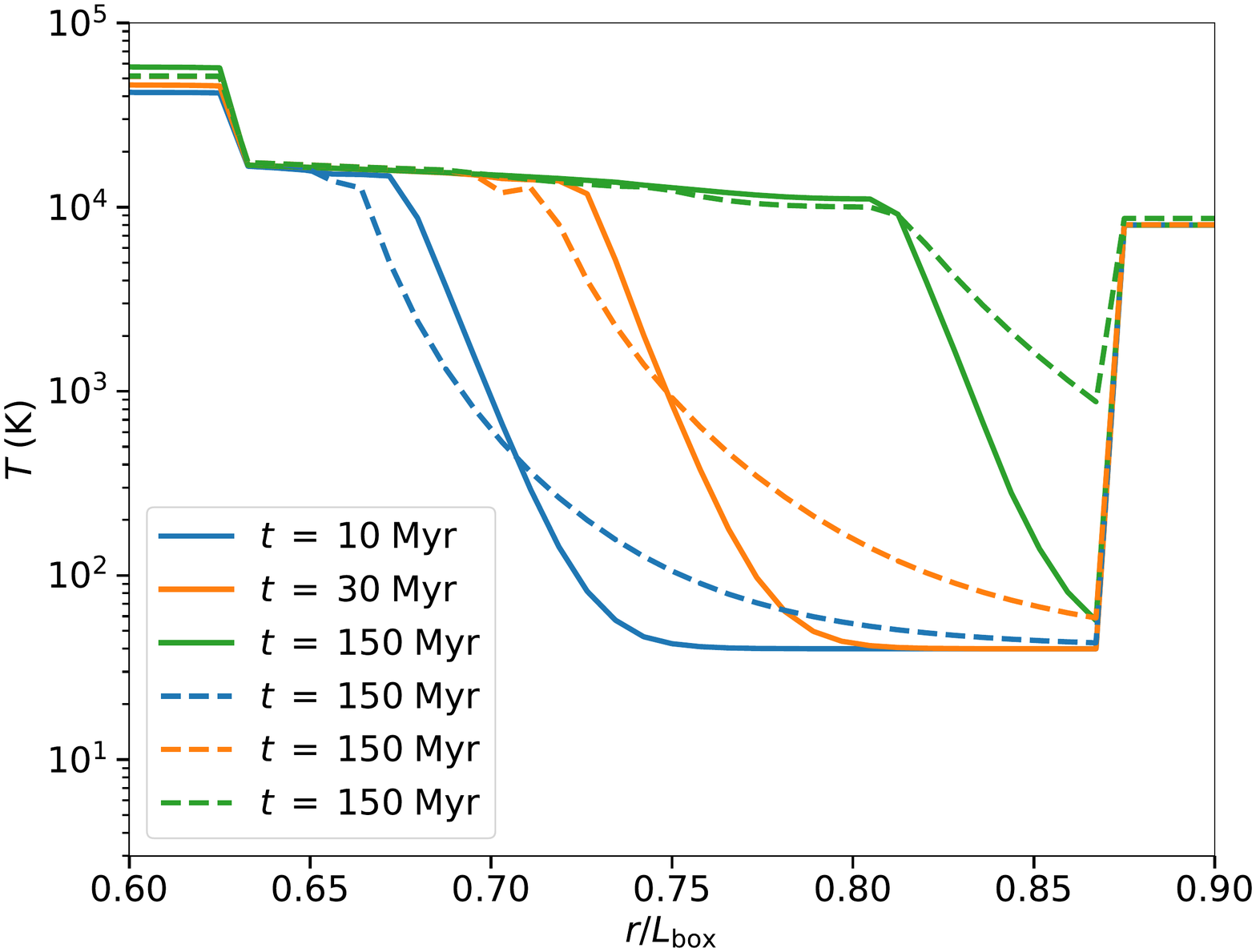}
  \caption{Test 3 - I-front trapping and formation of a shadow. ({\it Top and middle left.}) Slice plots of the neutral fraction at 1~Myr and 15~Myr into the simulation, respectively. ({\it Top and middle right.}) Slice plots of the temperature at 1~Myr and 15~Myr into the simulation. The shape of our shadow region expands with distance, a result of the point-like nature of the source of ionizing photons. Our results agree well with \citet{2011MNRAS.414.3458W}, a simulation which uses the same point-like configuration. Neutral fraction ({\it bottom-left}) and temperature ({\it bottom-right}) along a ray going through the center of the dense clump. The dashed lines represent the same plots calculated using $C^2$-Ray using data from RT06. We interpret the discrepancy in the right figure as the result of using a point source, as in our simulation, rather than a plan-parallel front. The diverging rays heat more effectively closer to the source (low $r$) and less effectively further from the source (high $r$).}
\label{fig:T3_0}
\end{figure*}
\begin{figure}[tbh]
\includegraphics[width=8.9cm]{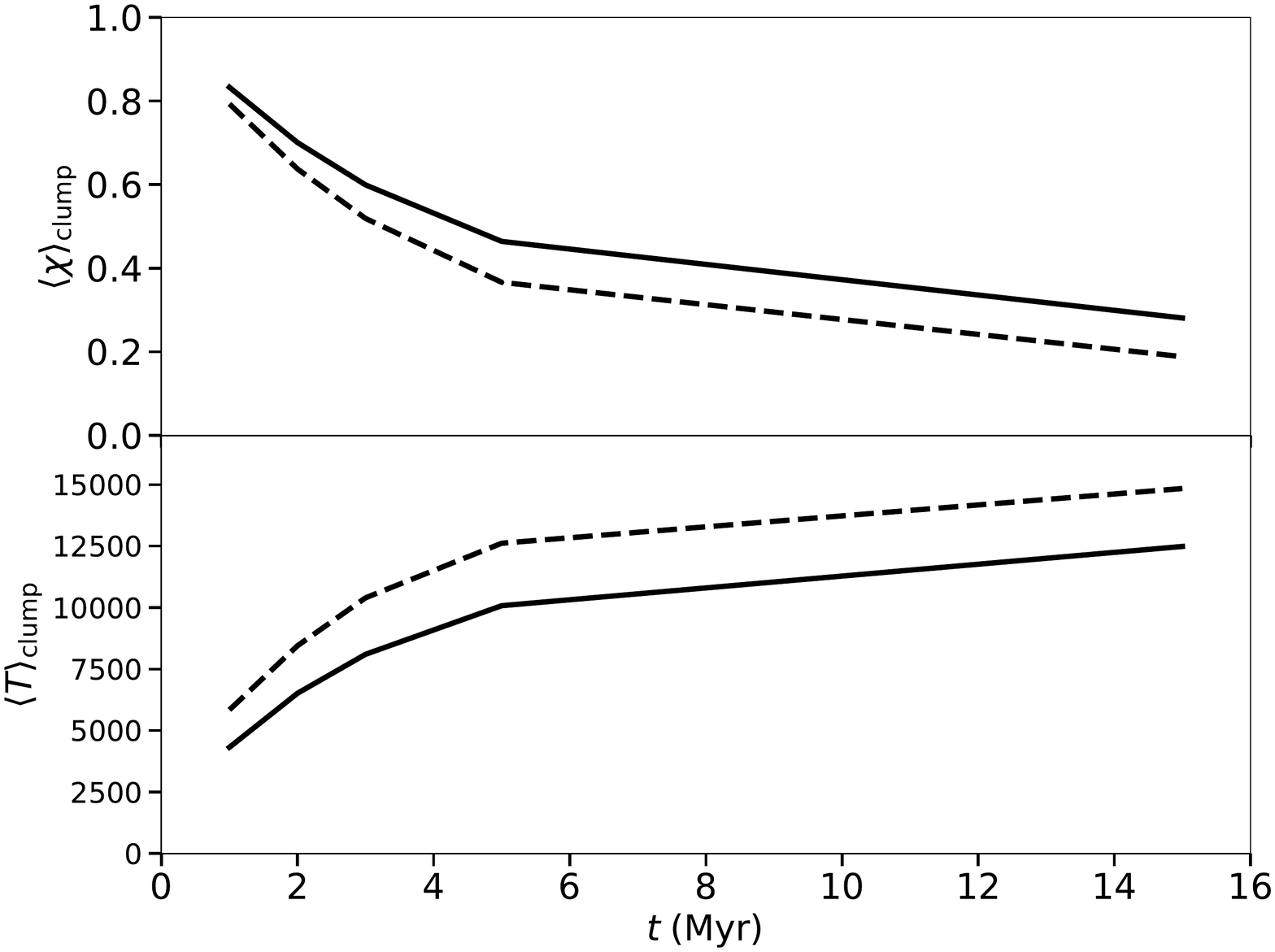}
\caption{Test 3 - I-front trapping and formation of a shadow. Average neutral fraction ({\it top}) and temperature ({\it bottom}) within the dense clump as a function of time during the simulation. The dashed lines represent the same results plotted using $C^2$-Ray data from RT06. We see that our ionization and heating are lower; we interpret this discrepancy as the result of adopting a point source of radiation rather than a plane-parallel front.}
\label{fig:T3_2}
\end{figure}

Test~3 in RT06 is designed to test the diffusivity and angular resolution of the radiative transfer code. In this test, a field of uniform radiation is projected towards a dense sphere of uniform hydrogen surrounded by a very thin medium. We plot the results of our model in Figure~\ref{fig:T3_0}. The I-front propagates at a constant velocity towards the clump until it reaches the surface, at which point the optically thick clump begins slowly absorbing the radiation. The lines of sight which pass through the sphere are trapped in the clump, causing a shadow to form behind the clump. The sharpness of the edge of the shadow is a measure of the diffusivity of the method. The edges of the sphere become ionized before the rest of the clump, causing the shadow to shrink; this allows to visually assess the angular resolution of the code. The test also tracks the rate at which the I-front progresses through the clump. 

The original test presented in RT06 is contained within a 6.6 kpc box with resolution $128^3$. The ambient medium has a density $n_{\rm out}=2\times10^{-4}\;{\rm cm}^{-3}$ and temperature $T_{\rm out, init}=8000$ K, while the clump has a density $n_{\rm clump}=0.04\;{\rm cm}^{-3}$ and temperature $T_{\rm clump, init}=40$ K. The clump has a radius 0.8 kpc and is centered at $(x_c,y_c,z_c)=(97,64,64)$ in grid units. The original test assumes plane-parallel radiation with flux $10^6\;{\rm s}^{-1}\;{\rm cm}^{-3}$. However, as in Test 0, we follow \cite{2011MNRAS.414.3458W} and replace the plane parallel radiation with a single point source of radiation opposite the clump, with $(x_c,y_c,z_c)=(0,64,64)$. The luminosity of the source is set at $S_0=3\times10^{51}\;{\rm photons}\;{\rm s}^{-1}$, so that the flux at the center of the clump matches the plane parallel flux of the original test. The simulation is evolved for 15~Myr, at which point RT06 find that the I-front is just past the center of the clump.

The top and middle left panels in Figure~\ref{fig:T3_0} show the neutral fraction in a slice through the center of the volume at $t=1$ Myr and $t=15$ Myr, respectively. The corresponding panels in the top and middle right show the temperature for the same slice and times. We notice immediately that the shadows in our simulation are opening with distance, a result of the relatively small distance of the point source from the clump, as opposed to the plane parallel radiation in the original test. We also see that the diffuse gas outside the clump is immediately ionized and photo-heated to the point of becoming optically thin. We also see that edge of the shadow is very sharp, with the neutral fraction going from $\sim1$ to $\sim10^{-4}$ in the space of $\sim 1-2$ cells.

The bottom panels in Figure~\ref{fig:T3_0} show the neutral fraction (bottom-left) and temperature (bottom-right) within the clump along the ray passing trough the center of the clump. The clump is centered at $r/L_{\rm box}=0.75$ and extends between $r/L_{\rm box}\sim0.6$ and $r/L_{\rm box}\sim0.90$. We see that the photo-heated gas outside the clump reaches values above $3\times10^4$~K, while the photo-heated gas inside the clump has a temperature below $2\times10^4$ K, a result of the increased cooling at higher gas density. We see that the neutral fraction rises and the temperature falls as we move through the I-front. The neutral fraction reaches $50\%$ at $r/L_{\rm box}\sim0.8$, a result consistent with RT06. For comparison we also plot the results of $C^2$-Ray as similarly colored dashed lines. While the neutral fractions agree well, there is more of a discrepancy in the temperatures. We believe this is a result of the ray divergence present with a point source that is absent in the case of plane parallel radiation. The diverging rays overheat the gas behind the front and underheat the gas in front of the front, an effect we see in our plot.

Finally, in Figure \ref{fig:T3_2} we plot the average neutral fraction and temperature inside the clump for our model (solid lines) against the results from $C^2$-Ray (dashed lines). While we see that our model both heats and ionizes the clump less effectively than $C^2$-Ray, we again believe  this is a result of the diverging nature of our radiation, which exposes a smaller portion of the sphere to the full radiation, at least initially. Despite this discrepancy, our results are still well within the spread of models shown in RT06.

\subsection{Test 4 - Multiple Sources in a Cosmic Density Field}


\begin{figure*}[tbhp]
\centering
\includegraphics[width=8.7cm]{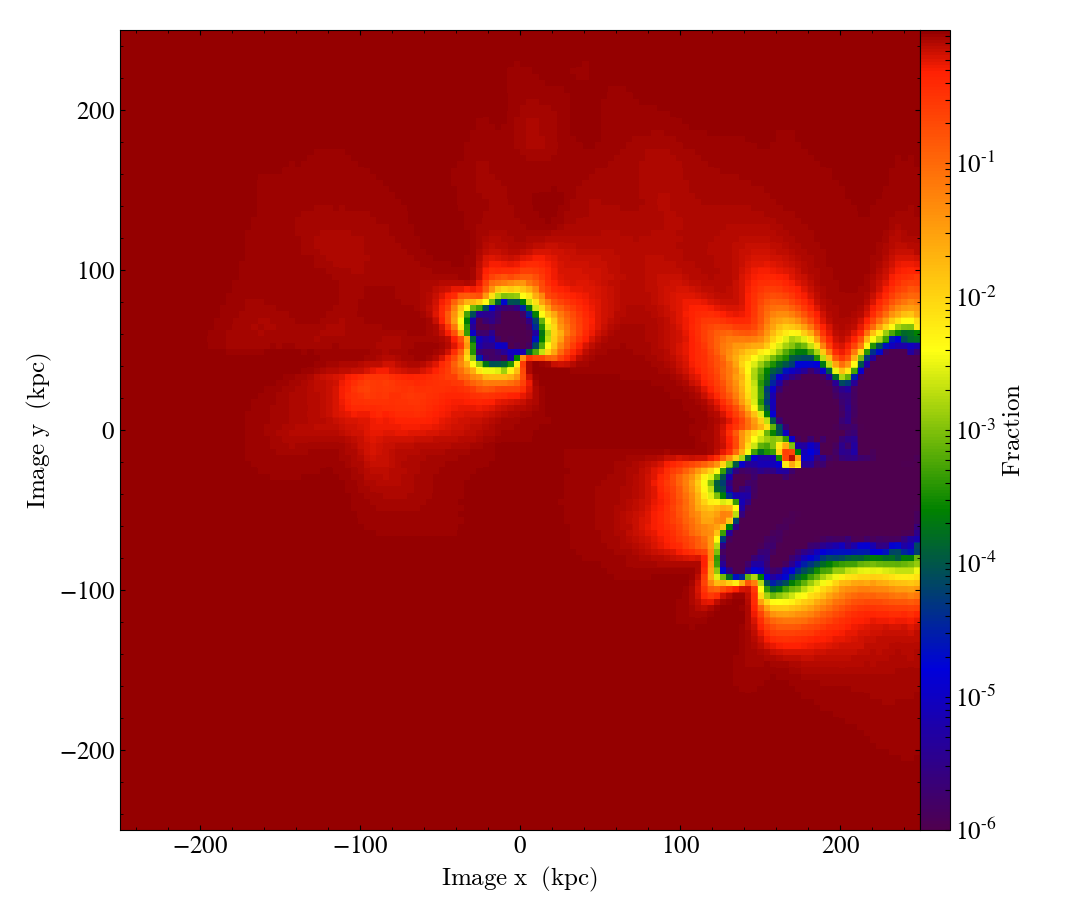}
\includegraphics[width=8.7cm]{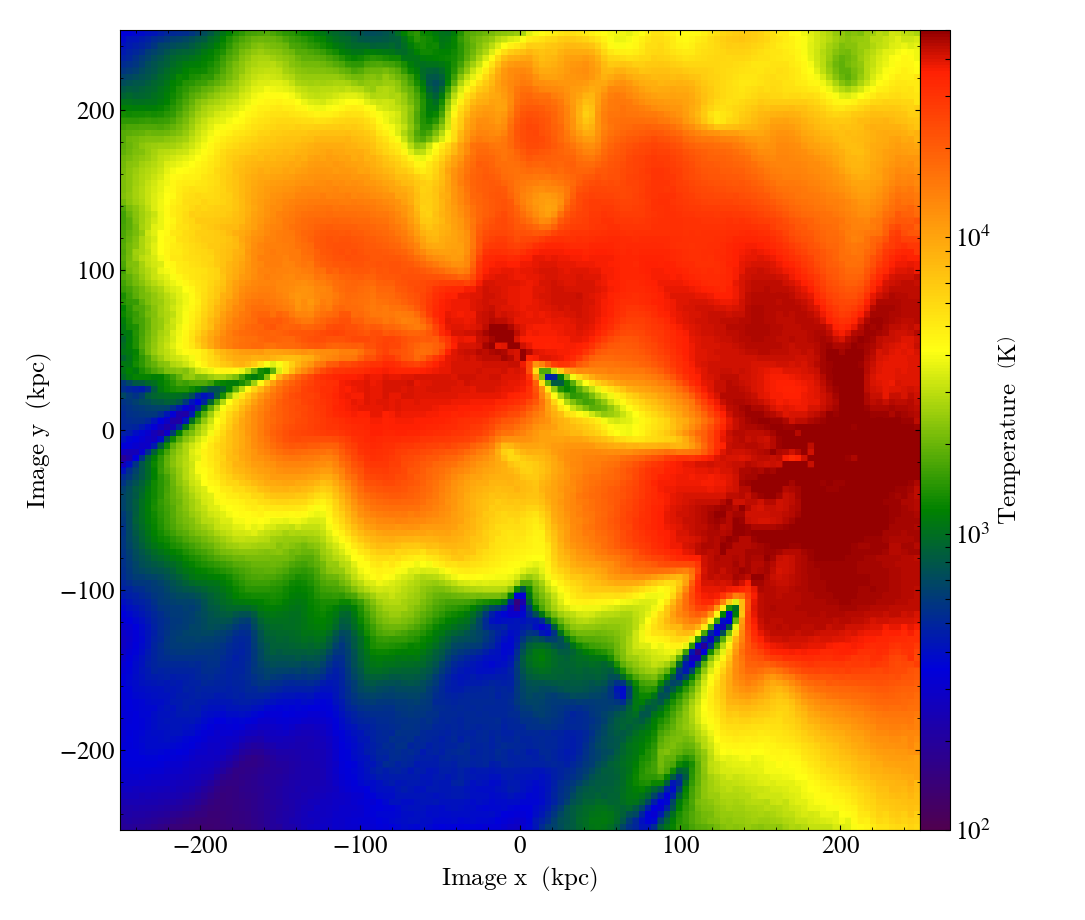}\\
\includegraphics[width=8.7cm]{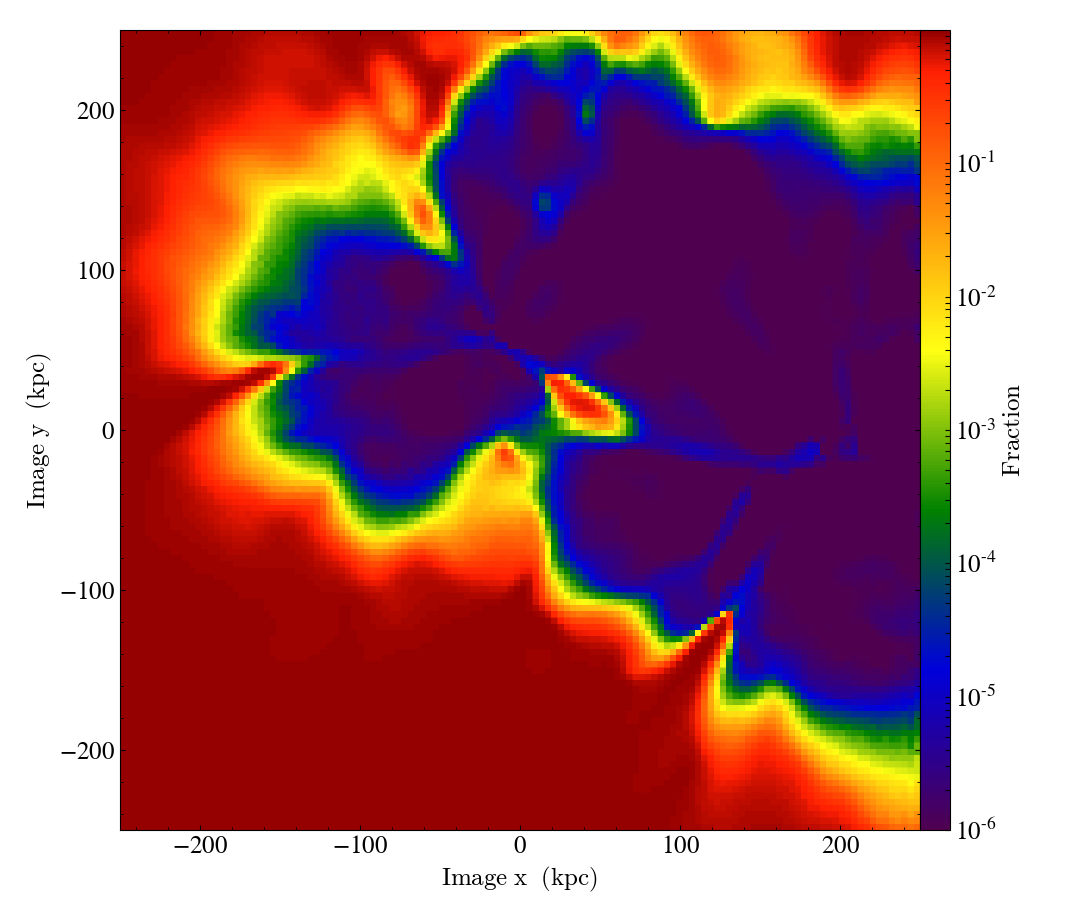}
\includegraphics[width=8.7cm]{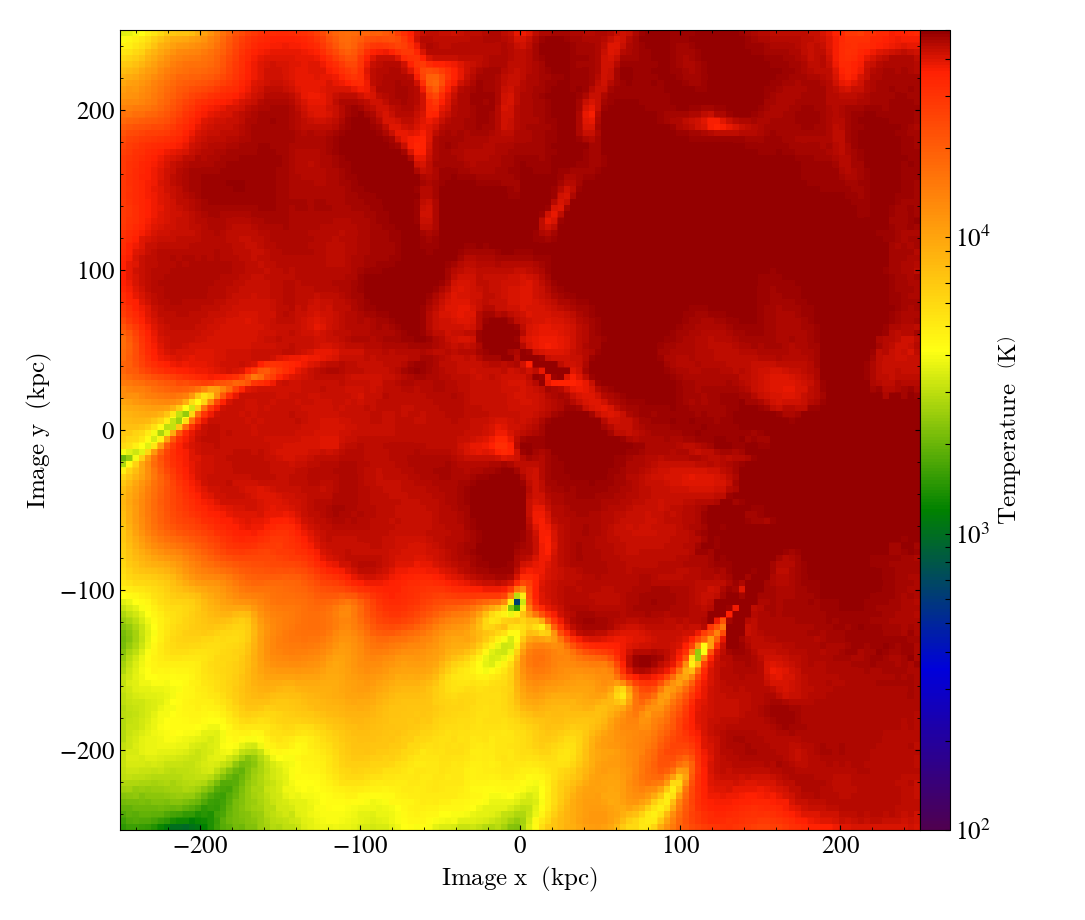}
\includegraphics[width=8.7cm]{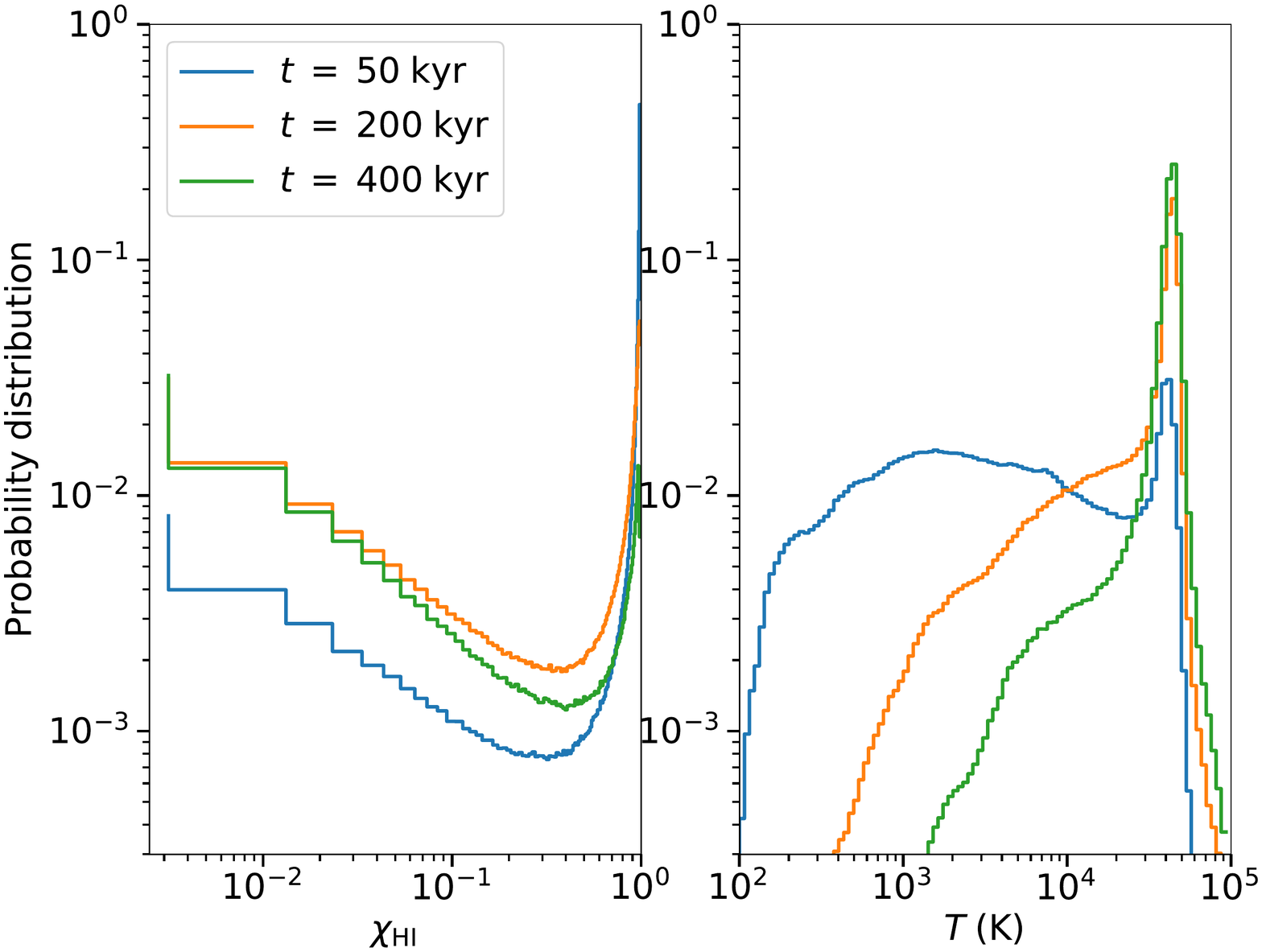}
\includegraphics[width=8.7cm]{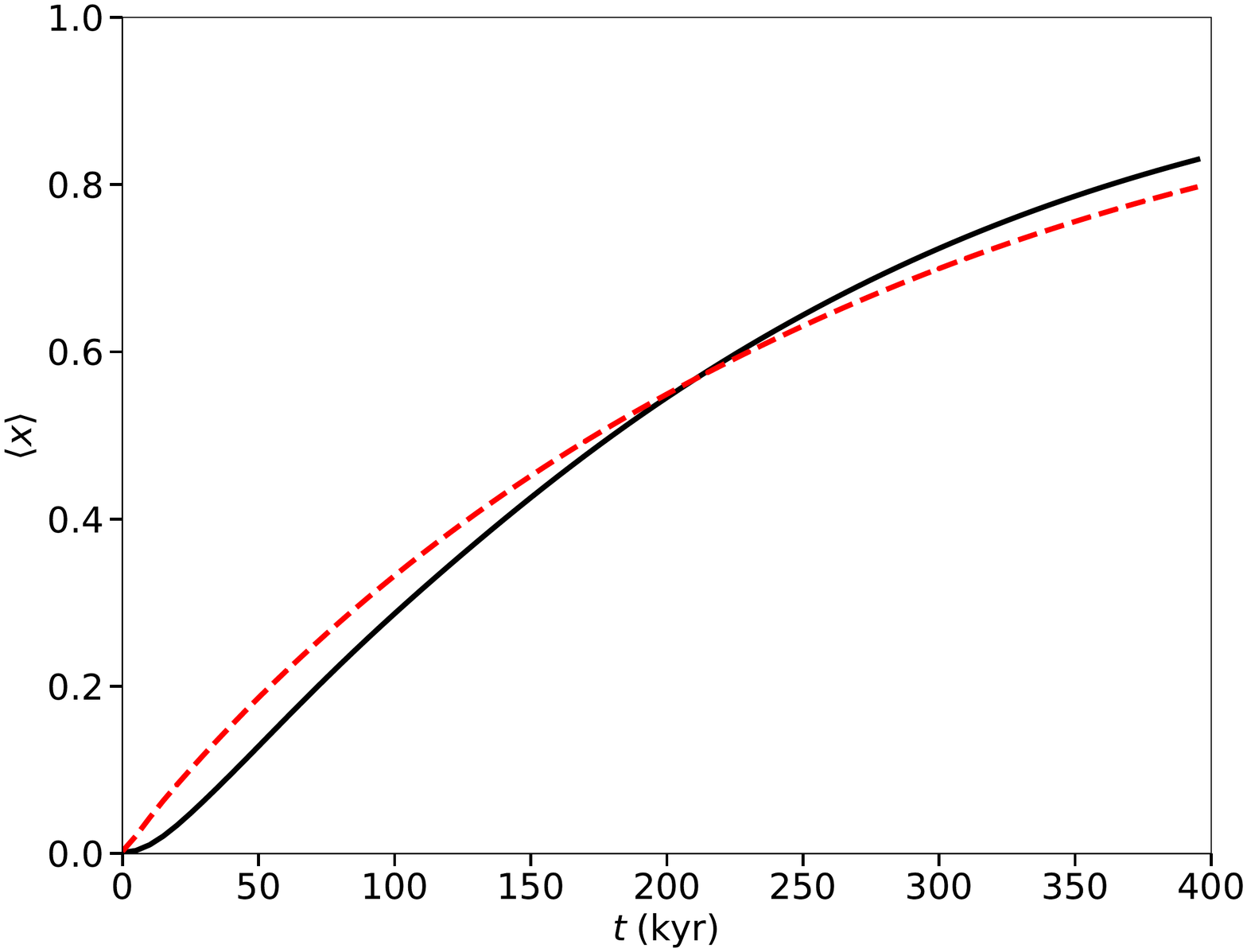}
\caption{Test 4 - Multiple sources in a cosmic density field. ({\it Top and middle left.}) Slice plots of the neutral fraction at 0.05 Myr and 0.2 Myr into the simulation, respectively. ({\it Top and middle right.}) Slice plots of the temperature at 0.05~Myr and 0.2~Myr into the simulation, respectively. These results are almost identical to those presented in \citet{2011MNRAS.414.3458W}, which uses a similar adaptive ray tracing scheme to ours. ({\it Bottom left.}) Histograms of neutral fraction (left) and temperature (right) at fixed times in the simulation. ({\it Bottom right}) Mass averaged (dashed line) and volume averaged (solid line) ionized fraction as a function of time throughout the simulation box. These results are in good agreement with the results presented in RT06.}
\label{fig:T4_0}
\end{figure*}

The final test in RT06 is a simple simulation of a cosmological density field. The simulation volume is cube with side $0.5\;h^{-1}$ cMpc at redshift $z=9$ and resolution $128^3$ (here $h=0.7$). The source of ionization is 16 point sources centered within the 16 most massive halos, and emit $f_{\gamma}=250$ ionizing photons per baryon with the same $T=10^5$ K blackbody spectrum used in tests 2 and 3. The sources are assumed to live for $t_s=3$ Myr, which is longer than the length of the simulation, so that they remain on for the entire simulation. The luminosity of each source is thus:
\begin{align*}
\dot{N}_{\gamma}=f_{\gamma}\frac{M\Omega_b}{\Omega_mm_Ht_s}.
\end{align*}

Here $M$ is the mass of the halo, $\Omega_b=0.043$, and $\Omega_m=0.27$. It is assumed that radiation leaving the box is lost, and the volume is tracked for a total period of 0.4 Myr. The density grid and halo position/luminosities are currently available from the Cosmological Radiative Transfer Comparison Project website.
\begin{figure*}[tbh]
\centering
\includegraphics[width=16.6cm]{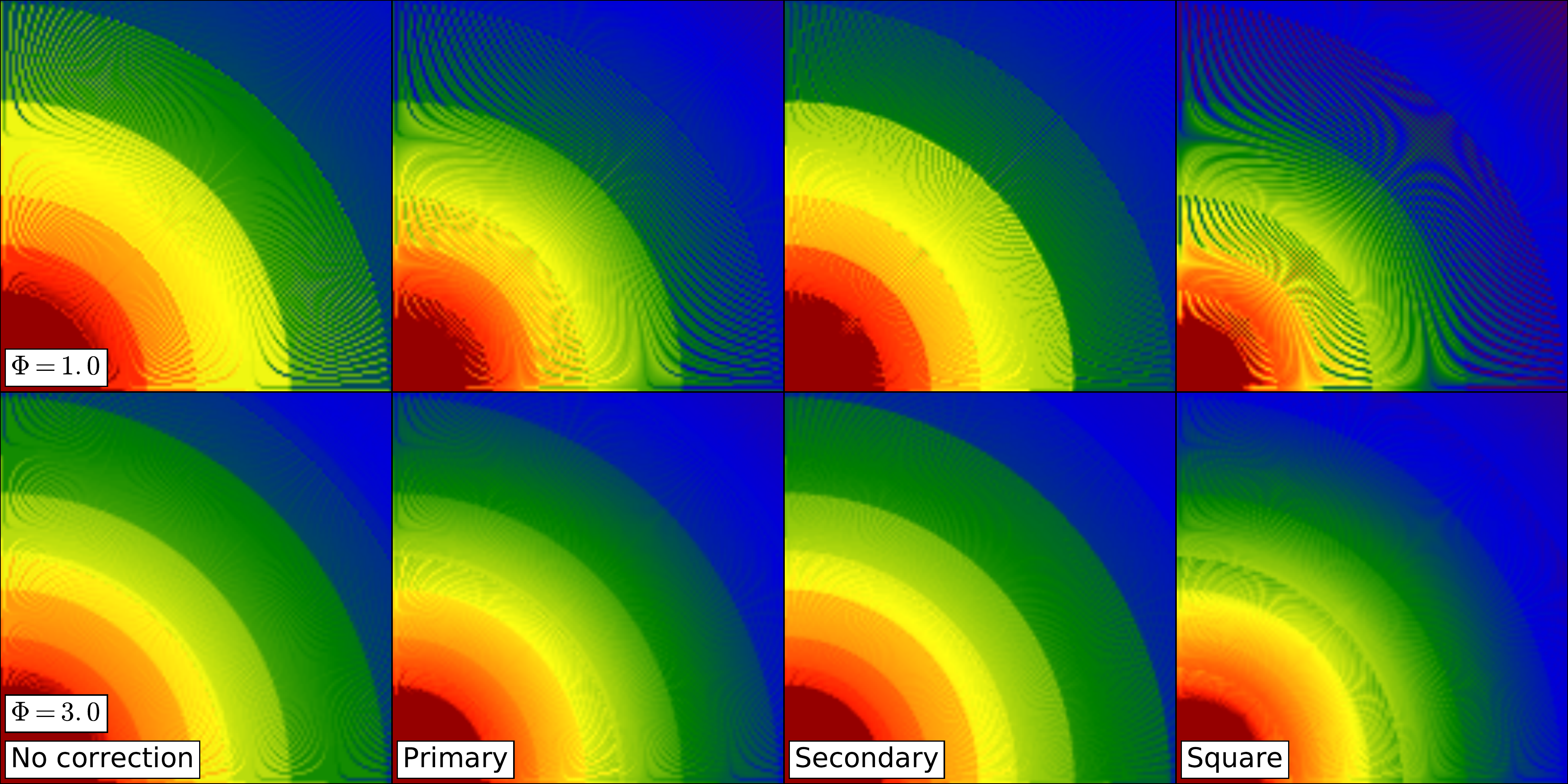}
\caption{This figure shows a selection of correction methods for two different ray splitting rates. The parameter $\Phi$ represents the number of rays per cell area, so that the bottom row have three times the splitting rate of those in the top row. The first column represents the method without correction factors. The second column illustrates the case with a primary correction, or the application of Equation~\ref{eq:correction} to the rate of ionization in every cell. The third column represents the case with primary and secondary correction, or the application of Equation~\ref{eq:correction2} to the cell which is closets to the cell face has the greatest overlap. The fourth column represents the case with only the primary correction with $f_c$ squared, as presented in \citet{2011MNRAS.414.3458W}. The low $\Phi=1$ plots were included to emphasize the difference between these configurations; clearly a higher value of $\Phi$ improves the accuracy for any configuration, but also increases the computational work for the simulation.}
\label{fig:mosaic}
\end{figure*}
In Figure~\ref{fig:T4_0} we show a selection of plots from this simulation. 
The top four figures are slices through the center of the simulation to allow for visual comparison between our results and the RT06 results. The ML and MR plots show slices through the neutral fraction grid at the $z=z_{\rm sim}/2$ plane of the simulation at $t=0.05$ Myr and $t=0.2$ Myr, respectively. The LL and LR plots show slices through the temperature grid at the $z=z_{\rm sim}/2$ plane of the simulation at $t=0.05$ Myr and $t=0.2$ Myr, respectively. The results are in good agreement, although the difference in colormap between our plots and those of RT06 may make visual comparison difficult. Our plots may also be visually compared to those of \cite{2011MNRAS.414.3458W}, which use a similar colormap and thus closely resemble our results.
\begin{figure}[hbt]
\includegraphics[width=8.6cm]{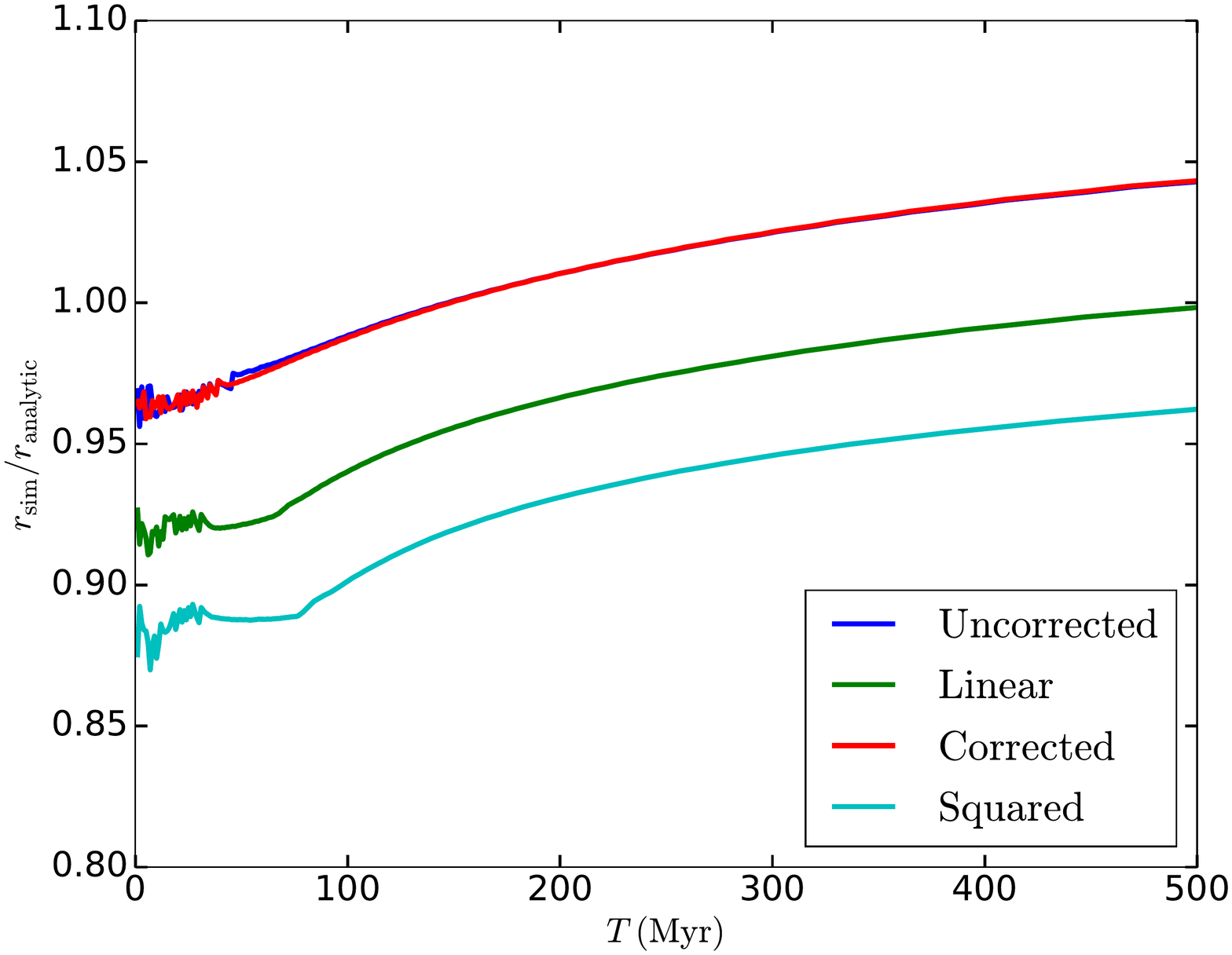}
\caption{Same as the bottom panel in Fig.~4 (left) but using different spherical correction factors. We see that the $f_c$ and $f_c^2$ corrections shrink the size of the ionized region by roughly 5\% and 10\%, respectively, while including a secondary correction to the $f_c$ correction restores the size of the region to the uncorrected size.}
\label{fig:rad_correction}
\end{figure}

The bottom-left panel in Figure~\ref{fig:T4_0} shows histogram of neutral fraction (left) and temperature (right) at times of 50, 200, and 400 kyr. These results are in good agreement with the codes presented in RT06. The bottom-right panel of Figure~\ref{fig:T4_0} shows the volume averaged ($\chi_v$, solid black line) and mass averaged ($\chi_m$, dashed red line) ionized fraction within the volume as a function of time. We see that the mass averaged ionized fraction is larger at early times, in agreement with the expectation of inside-out reionization within the RT06 simulation \citep{2000ApJ...542..535G,Miralda-Escude:2000,Sokasian:03}. These results are again in good agreement with those presented in RT06. 

\subsection{Methodology Test: Spherical Correction Factor}\label{ssec:methodtest}
\begin{figure*}[tbh]
\centering
\includegraphics[width=8.6cm]{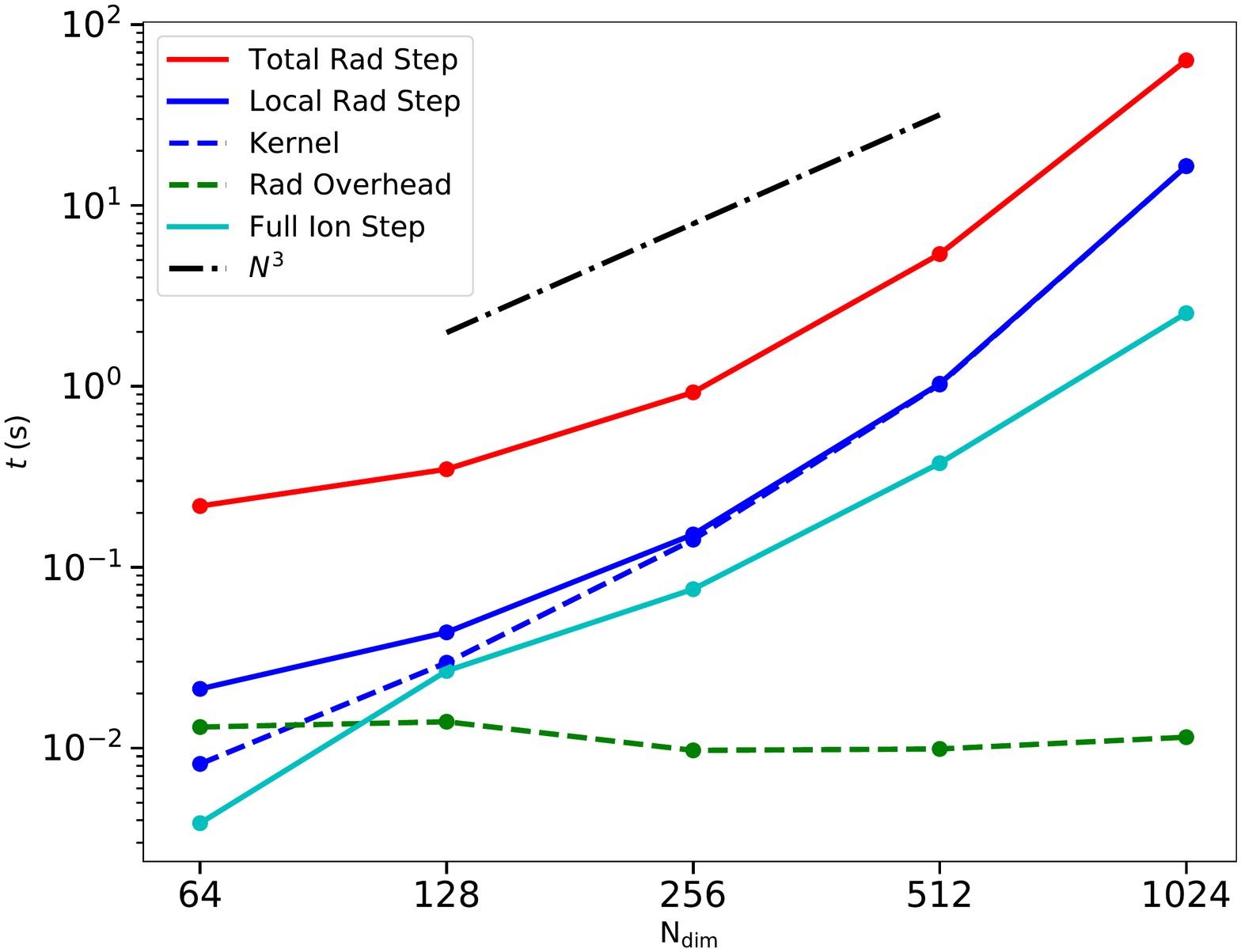}
\includegraphics[width=8.6cm]{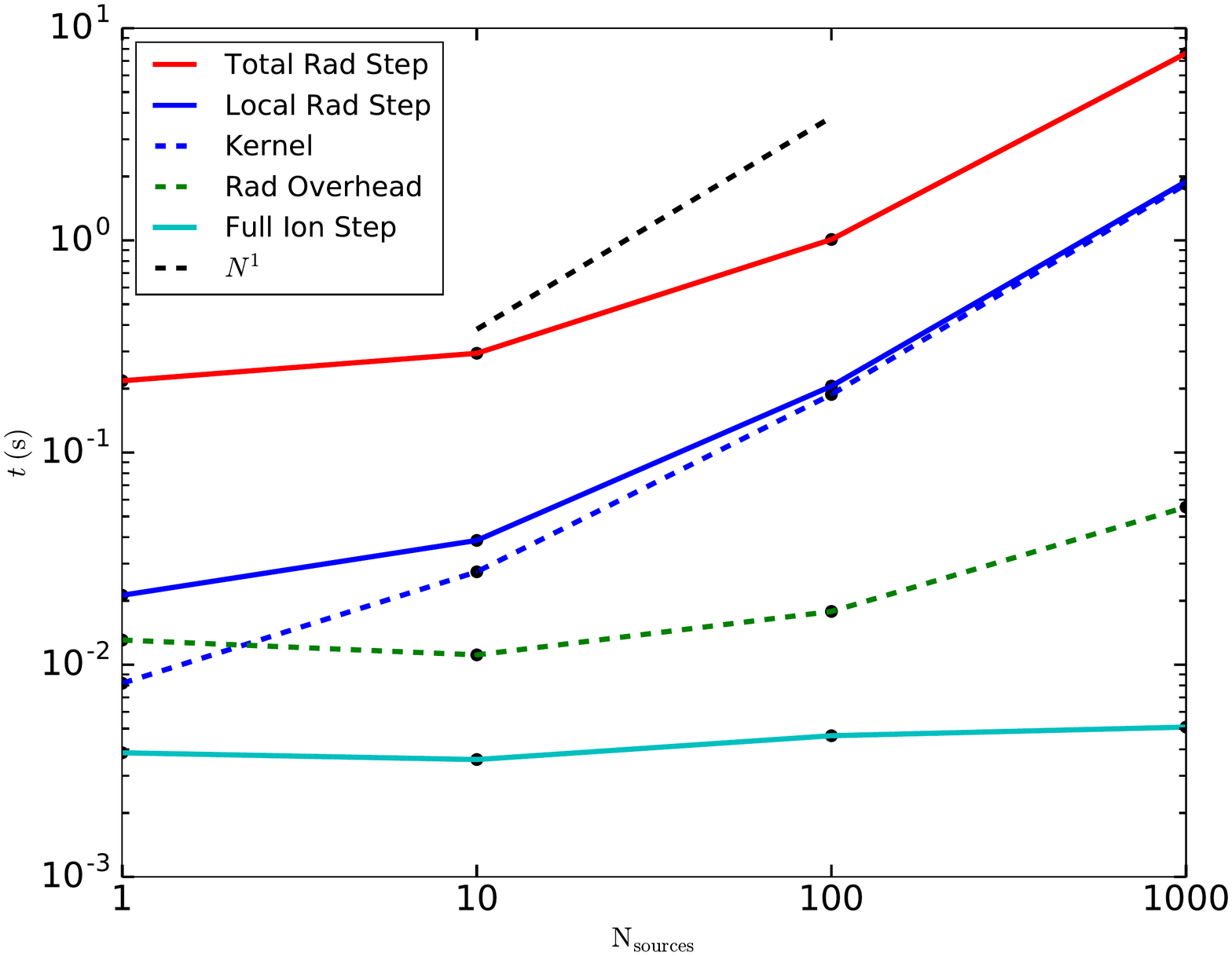}
\caption{Benchmarks for a Str{\"o}mgren sphere in a cubic grid. ({\it Left.}) The time taken to trace rays from a single source through the entire grid for grid sizes $64^3$ through $1024^3$. The time taken scales between $N^2$ and $N^3$. This is expected for this problem, as the number of rays scales with $N^2$ and the number of steps along each ray scales with $N$. ({\it Right.}) The time taken for a single radiative transfer step at grid size $64^3$ for a number of sources between $n=1$ and $n=1000$. For small source counts the per-step overhead can be seen, and as the number of sources increase, the time taken scales linearly, as expected.}
\label{fig:bench1}
\end{figure*}

The spherical correction factor defined in Section \ref{sec:correction} represents an attempt to approximate the true overlap between the cubic grid cells and the conic ray intersecting at arbitrary angles. The true correction factor depends on too many factors to calculate exactly, so any simple approximation will necessarily be somewhat flawed. In Figure~\ref{fig:mosaic} we plot a collection of slice plots from Test 1 using different choices for the power used in the correction factor formula. The top and bottom rows of this figure represent the same slices  with different choices of ray splitting rate ($\Phi=1$ and $\Phi=3$, respectively), while each column represents a different choice for the correction factor. The first column shows the case with no correction, and the artifacts are clearly visible. The second column shows the linear correction factor, and the second plot includes the secondary correction (which moves the excess radiation from a given cell to the adjacent cell which the ray is closest to). Finally, the last column shows the squared correction, as described in \cite{2011MNRAS.414.3458W}. We see that the linear power correction with the secondary correction results is the closest approximation of a true sphere.
The inclusion of this secondary correction also serves the purpose of compensating for the lost radiation due to the primary correction factor. In Figure~\ref{fig:rad_correction} we plot the comparison of the model radius to the analytic Str{\"o}mgrem radius for the same four models in Figure~\ref{fig:mosaic}. We see that the linear and square models reduce the radius of the model by factors of 5\% and 10\%, respectively. The correction factor, however, restores the accuracy of the model radius quite well.
We thus chose to use in our simulations the secondary correction in combination to the $f_c$ factor as it preserves photon conservation and removes azimuthal artifacts.

\section{Benchmarks}\label{sec:bench}

The way in which our code processes individual rays with GPUs is inherently parallel (CUDA kernel). For a single GPU, the number of cores over which this work is distributed is fixed by the specifications of the individual GPU. We thus choose to benchmark our code by measuring its performance on a simple test problem varying the number of source of radiation and the dimension of the computational grid. Our test problem is based on a test in \cite{2011MNRAS.414.3458W}. We place the ionizing source(s) at the center of a cubic $64^3$ grid of size 15~kpc. The medium is pure hydrogen of density $10^{-3}~{\rm cm}^{-3}$, and the ionizing source(s) have a monochromatic 17~eV spectrum with luminosity $5\times10^48$ photons/s. The simulation is run with the radiation-base adaptive time step (see Section~) for 250~Myr. For each simulations, we plot the time it takes for the entire ray tracing algorithm to run (Total Rad) as well as the time it takes for the algorithm to run on a single sub-volume (we divide the volume in 8 sub-volumes).

{\bf Dimension Scaling:} We measure how our code performs with varying grid size by performing the test simulation on grids by running the same simulation with grid resolutions of $64^3$, $128^3$, $256^3$, $512^3$, $1024^3$. The red line shows the time it takes for the full radiative transfer calculation, while the blue line shows the line shows the time it takes for a single sub-volume to complete its ray tracing calculation (the blue dashed line is the time it takes for the CUDA kernel to execute in isolation; this time is measured locally on the GPU, while the rest of the times are measured on the host node). The difference between these lines is the time it takes for the rays to be shared between the nodes and for the rays to propagate through the rest of the volume. We plot the difference between the total radiation calculation and the kernel execution times with the green line, as a measure of the MPI overhead during the radiative transfer calculation. Finally, we include the time taken for the ionization calculation with the cyan line. We see that the MPI overhead varies little with dimension, while the overall execution time scales with grid dimension as a power-law with slope between $N_{\rm dim}^2$ and $N_{\rm dim}^3$. The problem should theoretically scale with $N_{\rm dim}^3$: two powers of $N_{\rm dim}$ for the number of rays, and one power of $N_{\rm dim}$ for the number of steps along a ray of the same physical length. However, our GPU code doesn't perform as well when the number of rays is small, as the occupancy of the GPU processors is lower and some computational power is wasted.

{\bf Source Scaling:} We measure how our code performs with a varying number of sources by measuring how long the code takes to perform a single step of the calculation with $N$ sources at the same location, with the luminosity of each reduced by a factor of $N$, so that the final result should be the same. In the left plot of Figure~\ref{fig:bench1} we plot the resulting times for 1, 10, 100, and 1000 sources. We see that the code's overall time scales linearly with the number of sources, as anticipated.

\begin{figure}[tbh]
\centering
\includegraphics[width=8.6cm]{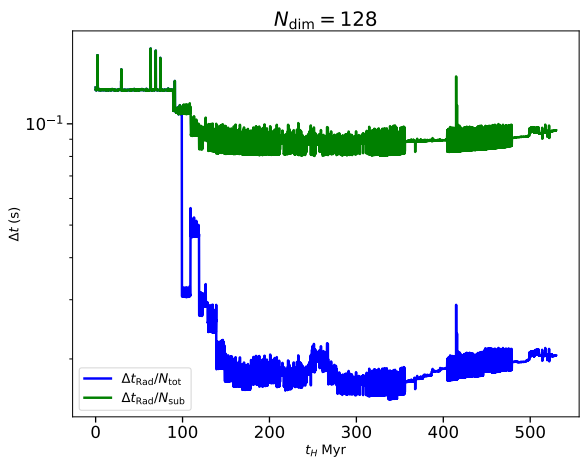}
\caption{Execution times (in seconds) for a $128^2$ simulation of cosmic reionization plotted as a function of simulation time (on 8 GPUs). The plot shows the time taken by a full radiative transfer calculation divided by the total number of sources (blue line) and divided by the maximum number of sources in any sub-volume (green line). The eight sub-volumes are processed in parallel, so as long as the sources are well distributed within the volume, the blue line should represent the mean time to execute the radiative transfer step per source at this resolution. We note that the times are unusually large at the beginning of the simulation, as the constant overhead dominates when the number of radiation sources is very small ({\it i.e.}, zero or one source) }
\label{fig:bench2}
\end{figure}

{\bf Cosmic Simulation Benchmark:} We also include a benchmark based on a full cosmic simulation run using our code on a $128^3$ grid divided evenly into 8 sub volumes, each $64^3$. In Figure \ref{fig:bench2} we plot the time taken for a complete ray tracing step divided by the total number of sources (blue) and by the number of sources in the most populated sub-volume (green). In such cosmic simulations, the bottlenecked is the sub-volume which takes the longest time to process, which is why we plot the time divided by the number of sources in the most populated volume. However, the since the GPU's process the rays independently, the time divided by the total number of sources is a better proxy for the time taken per step per source, as long as the sources are relatively evenly distributed between the sub volumes. In this simulation, when $x_e\sim0.2$, the simulation is able to fully process rays from $\sim 10^4$ sources on a $256^3$ grid in $\sim 20$ seconds. 

\section{Summary}\label{sec:summary}

We have described our implementation of the spatially adaptive ray tracing radiative transfer code ARC which is designed to take advantage of the extremely parallel processing power of the GPUs available in today's supercomputers. Our algorithm is based on the well known method presented in \cite{2002MNRAS.330L..53A}. We have presented the methodology of our code, as well as a novel approach to a correction method for ray tracing in a Cartesian grid. Our code is able to split a computational volume into sub-volumes, each of which is contained on an independent GPU linked by MPI, allowing our code to tackle very large problems avoiding the limitations related to the available memory on the GPU. We verified the accuracy of our method by performing a selection of tests presented in \cite{2006MNRAS.371.1057I}. Finally, we discussed a selection of benchmarks to demonstrate the speed and scaling of our code.

We believe that the unique characteristics of GPUs make them ideal for the computational problem of ray tracing. ARC already takes advantage relatively recent innovation such as CUDA aware MPI (only becoming available in 2013), but the optimization and speed up of our code is an ongoing process that will take advantage of new capabilities of GPUs as they become available. The speed, number of cores and the memory of GPUs has been growing rapidly over the years allowing us to tackle problems which were computationally unfeasible only a few years ago. 

The first application of ARC will be to simulate the reionization epoch using pre-computed dark matter simulations in a (10~cMpc)$^3$ volume, with sufficient resolution to capture minihalos with masses $>10^6$~M$_\odot$. These simulations can resolve the sites of formation of Population~III stars, which we will be able to model adopting the model in \citep{R2016}, and have a sufficiently large volume to capture the formation of galaxies observed in the Hubble ultra-deep fields. We will take advantage of the non-diffusive nature of the ray-tracing method to simulate realistic emission of ionizing radiation from the minihalos, including the short duration of the bursts of radiation and anisotropic emission expected if the dominant star formation mode is in compact star clusters \citep{R2002,KatzR:2013,KatzR:2014,RPG2016}. Based on our previous analytical study \citep{HartleyR:2016}, we believe that properly accounting for these effects will have a major impact on both the topology of reionization and the budget of ionizing photons necessary to reionize by redshift $z \sim 6.2$.


\subsection*{ACKNOWLEDGMENTS}
MR thank the National Science Foundation for support under the
Theoretical and Computational Astrophysics Network (TCAN) grant
AST1333514 and CDI grant CMMI1125285. Thanks to the anonymous
referee.

\bibliographystyle{./mn2e}
\bibliography{arc}

\label{lastpage}
\end{document}